\documentclass[aps,pra,preprint,superscriptaddress,showpacs]{revtex4-1}
\usepackage{graphicx}
\usepackage{dcolumn}
\usepackage{bm}
\begin{document}



\title{Theory of the Two-Particle Emission from Superfluid Fermi Gases in the BCS-BEC Crossover}


\author{Emiko Arahata}
\email{a1208702@rs.kagu.tus.ac.jp}
\affiliation{Department of Basic Science, The University of Tokyo, 3-8-1  Komaba, Meguro-ku, Tokyo,  153-8902, Japan}
\author{Tetsuro Nikuni}%
\affiliation{
Department physics, Faculty of science, Tokyo University of Science, \\
1-3 Kagurazaka, Shinjuku-ku, Tokyo 162-8601, Japan}%


\date{\today}

\begin{abstract}
We present a theory of the emission of fermion pairs from a superfluid Fermi gas induced by a photon absorption. In the solid state physics, this type of process is called double photo-emission (DPE). The spectrum of the induced two-particle current (or DPE current) provides a direct insight into the pair-correlation of condensate fermion pairs. We develop a general formalism for two-particle current induced by DPE by treating the coupling of two Fermi gases with the time-dependent perturbation theory. This formalism is used to calculate energy distributions of DPE current from the superfluid Fermi gas in the BCS-BEC crossover at $T=0$. 
We show that the DPE current has distinct contributions of the condensed pair components and uncorrelated pair states. 
We also calculate the angular dependence of DPE current in the BCS-BEC crossover.  The DPE current of the tightly-bound molecules in the BEC regime is found to be quite deferent from that of the weakly-bound Cooper pairs.
\end{abstract}

\pacs{67.85.-d,03.75.Ss, 05.30.Fk, 79.60.Ht}

\maketitle
\section{Introduction}
In recent experiments of ultracold atomic Fermi gases, the crossover from the BCS-type superfluid to the Bose-Einstein condensation (BEC) of tightly bound molecules including the unitary gas as an intermediate regime 
have been realized using a tunable pairing interaction associated with a Feshbach resonance \cite{E_Regal_PRL92,E_Zwierlein_PRL92,E_Zwierlein_NatureLondon435}. 
While a lot of studies are concerned with the thermodynamic properties of superfluidity as well as collective modes in ultracold atomic gases \cite{Rev_Giorgini_RMP2008}, the basic properties of condensate atom pairs is also interesting problem from the conceptual viewpoint. 
In particular, quasiparticle excitations were studied by using the momentum-resolved photoemission-type spectroscopy, in which atoms are transferred to the third empty atomic state by rf pulse\cite{E_photo_Nature}. This powerful technique, which is an analog
of the angle-resolved photoemission spectroscopy (ARPES) in solid state physics, allows one to measure microscopic properties of a cold Fermi gas
in the crossover region. 
The photoemission-type spectroscopy has been theoretically addressed
in the literature of cold Fermi gas \cite{T_photo_Tsuchiya,T_photo_Torma}. 
The experimental results are in qualitative agreement with the theoretical calculation. 

In this paper, we consider an alternative approach to study the basic properties of condensate atom pairs in a superfluid Fermi gas.
The main purpose of this paper is to distinguish between cooper pairs and tightly bound molecules in the BCS-BEC crossover from the emission of Fermi condensate pair.
In solid state physics, emission of election pair induced by photon absorption is called the double photoemission (DPE) and is used to measured two-particle spectra that provide direct insight into the energy and the angular dependence of the pair-correlation functions \cite{T_Fominykh_SolidStateCommun,T_Kouzakov_PRL}.
Many experiments have observed DPE spectroscopy from conventional and unconventional superconducting samples \cite{ET_Fominykh_PRL,E_Herrmann_PRL,E_Christianson_Nature}.
In view of the advances in experimental techniques in ultracold atomic gases, an analogous experiment on Fermi atomic superfluid may be expected to become available.
In this paper, we provides a general theory of two-particle current induced by DPE (DPE current) from superfluid Fermi gases \cite{B_Mahan, T_Luxat_PRA2002}.
We note that the previous studies of the photoemission-type spectroscopy \cite{T_photo_Tsuchiya,T_photo_Torma} essentially deals with the situation where a single atom is emitted from the system by absorbing a photon. In the present paper, we consider a possibility of emission of a pair of atoms by photon absorption. As an illustration, we consider the BCS-BEC crossover at $T=0$ within the framework of BCS-Leggett's theory. We calculate the energy dependence of DPE current, and explicitly show that the contribution of uncorrelated pair states and that of condensed pairs are clearly separated.
We will also show that the angular dependence of DPE current, which indicate the contribution of condensed pairs of DPE current see clear compared to single-photoemission current.

In Sec. II, we will derive a general expression for DPE current of atoms tunneling between two atomic gases coupled through an external field. 
By treating the coupling between two fermi gases as perturbation, we employ time-dependent perturbation theory up to forth order in order to obtain non-vanishing contribution to two-particle current.    
We will then introduce a two-particle spectral function describing DPE current. 

In Sec. III, we will calculate the two-particle spectral function in the BCS-BEC crossover at $T=0$ using Leggett's theory \cite{T_Luxat_PRA2002} based on the mean-field treatment.  

In Sec. IV, we will show the calculations of DPE current as a function of the energy transfer from coupling field, and discuss the separate contributions of the condensed pair components and uncorrelated pair states. We will also show the angular distributions of DPE current and discuss the possibility to distinguish between Cooper pairs and molecules form the calculations of DPE current. 

For comparison, we will also show the single-particle current in the BCS-BEC crossover in Sec. V including the higher-order process involving two-particle tunneling. We will show the contribution of uncorrelated pair states is much lager than that of condensed pairs, although there appears a small peak from contribution of condensed pairs.

\section{General Formalism for Two-Particle Current of a two-component Fermi gas}
\begin{figure}[htbp]
\includegraphics[width=50mm]{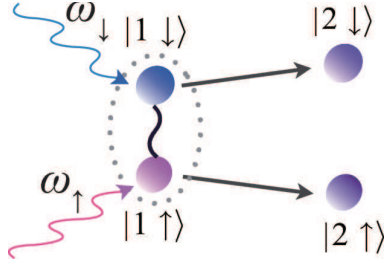}
\caption{\label{fig:Fig_tra} (Color online) The illustration for emission of fermion pairs from two-particle induced by a photon absorption from a superfluid Fermi gas. }
\end{figure}

In this section, we present a formalism for two-particle current induced by DPE by treating the coupling of two Fermi gases with the time-dependent perturbation theory. 
In Fig.~\ref{fig:Fig_tra}, we illustrate the emission of fermion pairs induced by a photon absorption from a superfluid Fermi gas.
We assume that atoms are initially in state 1 and transfered to the state 2 when a coupling interaction is switched on. 
The tunneling perturbation $V(t)$ couples the two many-body systems together by introducing a mechanism by which an atom can tunnel between the two systems. This perturbation has terms that create an atom in one system while destroying
an atom in the other system and vice versa.

The total Hamiltonian for a two-component Fermi gas is given by
\begin{eqnarray}
\hat H=\hat H_1+\hat H_2 + \hat V(t)\equiv \hat H_0+ \hat V(t),
\label{eq:Hami}
\end{eqnarray}
where $\hat H_1$ describes the initial state of interest, and
$\hat H_2$ describes the final state that is coupled to the state 1
through the tunneling Hamiltonian $\hat V(t)$ given by 
by
\begin{equation}
\hat V(t)=\sum_{\sigma}\sum_{{\bf k}}
e^{\eta t}
\left(\gamma_{\sigma}e^{-i \omega_{\sigma}t}
\hat b^{\dagger}_{{\bf k}+{\bf q}_{\sigma}\sigma}
\hat c_{{\bf k}\sigma}+{\rm H.c.}\right).
\label{double_eq_2}
\end{equation}
Here $\hat c_{{\bf k}\sigma}$ and $\hat b_{{\bf k}\sigma}$ are creation operators 
of the state 1 and 2.
$\omega_{\sigma}$ and ${\bf q}_{\sigma}$ are the effective energy transfer and the momentum
transfer from the coupling fields, and $\gamma_{\sigma}$ is the coupling strength.
We have also introduced a factor $e^{\eta t}~~ (\eta>0)$ that models the adiabatic switching on of the interaction at
$t\to-\infty$.
\subsection{Two-Particle Current}
Let us define the two-particle density for the state 2  as
\begin{equation}
n({\bf k}_1\uparrow,{\bf k}_2\downarrow,t)
\equiv
\langle \hat b^{\dagger}_{ {\bf k}_1\uparrow}
\hat b^{\dagger}_{ {\bf k}_2\downarrow}
\hat b_{ {\bf k}_2\downarrow}
\hat b_{ {\bf k}_1\uparrow}\rangle_t
\equiv
\langle \hat n({\bf k}_1\uparrow,{\bf k}_2\downarrow) \rangle_t,
\end{equation}
where \begin{equation} \hat n({\bf k}_1\uparrow,{\bf k}_2\downarrow)
\equiv 
\hat b^{\dagger}_{ {\bf k}_1\uparrow}
\hat b^{\dagger}_{ {\bf k}_2\downarrow}
\hat b_{ {\bf k}_2\downarrow}
\hat b_{ {\bf k}_1\uparrow}.
\end{equation}
This two-particles density describes the density of fermion pairs whose momentum ${\bf k}_1$ and ${\bf k}_2$. 

Two-particle current is defined by time derivative of the two-particle
density
\begin{equation}
J({\bf k}_1\uparrow,{\bf k}_2\downarrow)=\frac{d}{dt}n({\bf k}_1\uparrow,
{\bf k}_2\downarrow)=\frac{i}{\hbar}\left\langle \left[
\hat n({\bf k}_1\uparrow,{\bf k}_2\downarrow,t), \hat H
\right] \right\rangle_t.
\label{J_5}
\end{equation}
This two-particle current describe the number of atom pairs emitted per unit time.
One can easily show that $\hat H_1$ and $\hat H_2$ make no contribution to the
commutator in the r.h.s. of Eq.~(\ref{J_5}) and thus
\begin{equation}
J({\bf k}_1\uparrow,{\bf k}_2\downarrow)=
\frac{i}{\hbar}\left\langle \left[
\hat n({\bf k}_1\uparrow,{\bf k}_2\downarrow), \hat V(t)
\right] \right\rangle_t.
\end{equation}
Using the expression (\ref{double_eq_2}) for the tunneling Hamiltonian, we find that the two-particle current is given by  
\begin{equation}
J(t)=e^{\eta t}\frac{i}{\hbar}
\left[
\gamma_{\uparrow}e^{-i\omega_{\uparrow}t}
\langle 
\hat F_{\uparrow}({\bf k}_1,{\bf k}_2,{\bf k}_2,{\bf k}_1-{\bf q}_{\uparrow})
\rangle_t+
\gamma_{\downarrow}e^{-i\omega_{\downarrow}t}
\langle
\hat F_{\downarrow}({\bf k}_2,{\bf k}_1,{\bf k}_1,{\bf k}_2-{\bf q}_{\downarrow})\rangle_t-{\rm c.c.}
\right],
\end{equation}
where we have introduced the following four field correlation function:
\begin{equation}
\hat F_{\uparrow}({\bf k}_1,{\bf k}_2,{\bf k}_3,{\bf k}_4)\equiv
\hat b^{\dagger}_{ {\bf k}_1\uparrow}\hat b^{\dagger}_{{\bf k}_2\downarrow }
\hat b_{{\bf k}_3\downarrow }\hat c_{{\bf k}_4\uparrow },~~~
\hat F_{\downarrow}({\bf k}_1,{\bf k}_2,{\bf k}_3,{\bf k}_4)\equiv
\hat b^{\dagger}_{ {\bf k}_1\downarrow}\hat b^{\dagger}_{{\bf k}_2\uparrow }
\hat b_{{\bf k}_3\uparrow }\hat c_{{\bf k}_4\downarrow }.
\end{equation}
Therefore, the two-particle current is expressed in terms of the correlation functions  $\langle\hat F_{\uparrow}\rangle$ and $\langle\hat F_{\downarrow}\rangle$. In the next section, we employ the time-dependent perturbation theory to derive expressions for these
correlation functions.
\subsection{General formalism of time evolution}

In general, nonequiliblium statical average of an arbitrary operator $\hat O$ is given by    

\begin{equation}
\langle \hat O \rangle={\rm Tr}\{\hat \rho(t)\hat O\}
={\rm Tr}\{\hat U(t,t_0)\hat \rho_0(t)\hat U^\dagger(t,t_0)\hat O\}={\rm Tr}\{\hat \rho(t_0)\hat O_H(t)\},
\end{equation}
where $\hat \rho(t)$ is the nonequilibrium statistical density operator, $t_0$ is the initial time, and $U(t,t_0)$ is the time evolution operator,
\begin{equation}
\hat U(t,t_0)={\mathcal T}\exp\left[-\frac{i}{\hbar }\int_{t_0}^{t} dt' \hat H(t')\right] ,
\end{equation}
$\mathcal{ T}$ being time-ordering operator.
$\hat O_H$ is the Heisenberg operator denied by 
\begin{equation}
\hat O_H(t)=\hat U^{\dagger}(t,t_0)\hat O\hat U(t,t_0).
\end{equation}
In order to perform perturbative expansion in the tunneling Hamiltonian $\hat V(t)$, we introduce the Heisenberg operator with respect to $H_0$ as
\begin{equation}
 \hat O_{H_0}(t) \equiv e^{iH_0(t-t_0)/\hbar} \hat O e^{-i\hat H_0(t-t_0)/\hbar}.
\end{equation}
Then, $\hat O_H$ and $\hat O_{H_0}$ are related though the unitary transformation  

\begin{equation}
\hat O_H(t)=\hat \mathcal{U}^{\dagger}(t,t_0)\hat O_{H_0}(t)\hat \mathcal{U}(t,t_0).
\end{equation}
where 
\begin{equation}
\hat \mathcal{U}(t,t_0)={\mathcal T}\exp\left[-\frac{i}{\hbar }\int_{t_0}^{t} dt' \hat V_{H_0}(t')\right] ,
\end{equation}
Therefore, we find
\begin{equation}
\langle \hat O \rangle_t=
\left\langle {\mathcal T}\exp\left[-\frac{i}{\hbar }\int_{t_0}^{t} dt' \hat V_{H_0}(t')\right] \hat O_{H_0}(t_0) \tilde{\mathcal T}\exp\left[\frac{i}{\hbar }\int_{t_0}^{t} dt' \hat V_{H_0}(t')\right]\right\rangle_{t_0}.
\end{equation}
where $\tilde{\mathcal T}$ is the anti-chronological time-ordering operator. 

The above expressions are conveniently denoted by using the Keldysh contour-time-path description \cite{B_Rammer}.
In general, the nonequilibrium expectation value of the physical quantity can be written as (we omit the subscript $H$ for simplicity)
\begin{equation}
\langle \hat O\rangle_t=\left\langle {\mathcal T}_c \left[\exp\left(-\frac{i}{\hbar}\oint_{t_0}^{t_0}dt' \hat V(t')\right) \hat O(t)\right]\right\rangle_{t_0},
\label{eq17}
\end{equation}
where ${\mathcal T}_c$ is a contour-ordering operator and 
\begin{equation}
\oint_{t_0}^{t_0}=\int_{t_0}^t+\int_{t}^{t_0}=\int_{t_0}^t-\int_{t_0}^{t}
\end{equation}
stands for the integral along the contour.
Expanding (\ref{eq17}) in the perturbation Hamiltonian $\hat V$, we obtain the following expansion 
  \begin{eqnarray}
  \langle \hat O \rangle_t=\sum_{n=0}^\infty  \langle \hat O \rangle_t^{(n)},
  \end{eqnarray}
  where the $n$-th order contribution is expressed as
    \begin{eqnarray}
   \langle \hat O \rangle_t^{(n)}\equiv \frac{1}{n!}\left(\frac{i}{\hbar}\right)^n 
   \oint_{t_0}^{t_0}dt_1\cdots \oint_{t_0}^{t_0} dt_n 
   \left\langle {\mathcal T}\left[V(t_1)\cdots \hat V(t_n)\hat O(t)\right]\right\rangle_{t_0}.
   \label{eq18}
  \end{eqnarray}
\subsection{General Expression for Two-Particle Current}

We now consider the quantity $\langle F_a \rangle$ defined in Sec.~II~A.
Under the assumption that the system 1 and 2 are uncoupled in the absence of the coupling Hamiltonian $V$, the first and second order contributions in the perturbative expansion (\ref{eq18}) vanish.  
We thus left with the third-order term as the lowest-order non-vanishing contribution. 
The expression for $\langle \hat F_{\sigma}\rangle$ is then given by
\begin{eqnarray}
&&\langle \hat F_{\uparrow}({\bf k}_1,{\bf k}_2,{\bf k}_2,{\bf k}_1-{\bf q}_{\uparrow}) \rangle_t^{(3)}=
\left(-\frac{i}{\hbar}\right)^3
\gamma_{\uparrow}^*|\gamma_{\downarrow}|^2
e^{i \omega_{\uparrow}t}
\int_{t_0}^{0} dt_1 \int_{t_0}^{0} dt_2 \int_{t_0}^{0} dt_3
\nonumber \\
&&
e^{i(\omega_{\uparrow}-\epsilon_{{\bf k}_1\uparrow}/\hbar-i\eta)t_1}
e^{i(\omega_{\downarrow}-\epsilon_{{\bf k}_2\downarrow}/\hbar-i\eta)t_2}
e^{-i(\omega_{\downarrow}-\epsilon_{{\bf k}_2\downarrow}/\hbar+i\eta)t_3}
 \nonumber \\
&&~~~~\times
\Bigl\langle \tilde{\mathcal T}\Bigl[\hat c^{\dagger}_{{\bf k}_2-{\bf q}_{\downarrow}\downarrow}(t_2)
\hat c^{\dagger}_{{\bf k}_1-{\bf q}_{\uparrow}\uparrow}(t_1)\Bigr] 
\hat c_{{\bf k}_1-{\bf q}_{\uparrow}\uparrow }(0)\hat c_{{\bf k}_2-{\bf q}_{\downarrow}\downarrow}(t_3) \Bigr\rangle_{t_0},
\label{eq45a}
\\
&&\langle \hat F_{\downarrow}({\bf k}_2,{\bf k}_1,{\bf k}_1,{\bf k}_2-{\bf q}_{\downarrow}) \rangle_t^{(3)}=
\left(-\frac{i}{\hbar}\right)^3
\gamma_{\downarrow}^*|\gamma_{\uparrow}|^2
e^{i \omega_{\downarrow}t}
\int_{t_0}^{0} dt_1 \int_{t_0}^{0} dt_2 \int_{t_0}^{0} dt_3 
\nonumber \\
&&\times
e^{i(\omega_{\downarrow}-\epsilon_{{\bf k}_2\downarrow}/\hbar-i\eta)t_1}
e^{i(\omega_{\uparrow}-\epsilon_{{\bf k}_1\uparrow}/\hbar-i\eta)t_2}
e^{-i(\omega_{\uparrow}-\epsilon_{{\bf k}_1\uparrow}/\hbar+i\eta)t_3}
 \nonumber \\
&&~~~~\times
\Bigl\langle \tilde{\mathcal T} \Bigl[\hat c^{\dagger}_{{\bf k}_1-{\bf q}_{\uparrow}\uparrow}(t_2)
\hat c^{\dagger}_{1{\bf k}_2-{\bf q}_{\downarrow}\downarrow}(t_1)\Bigr] 
\hat c_{{\bf k}_1-{\bf q}_{\downarrow}\downarrow }(0)\hat c_{{\bf k}_1-{\bf q}_{\uparrow}\uparrow}(t_3) \Bigr\rangle_{t_0},
\label{eq45b}
\end{eqnarray}
As we noted before, we assume that $\eta|t|<<1$.

Let us define the two-particle correlation functions by
\begin{eqnarray}
&&iG_{\uparrow}({\bf k}_1',{\bf k}_2',t_1,t_2,t_3)\equiv\Theta(-t_1)\Theta(-t_2)\Theta(-t_3)
\Bigl\langle \tilde{\mathcal T}\Bigl[\hat c^{\dagger}_{{\bf k}'_2\downarrow}(t_2)
\hat c^{\dagger}_{{\bf k}'_1\uparrow}(t_1)\Bigr] 
\hat c_{{\bf k}_1'\uparrow }(0)\hat c_{{\bf k}_2'\downarrow}(t_3) \Bigr\rangle_{t_0},\label{eq_22}
\\
&&iG_{\downarrow}({\bf k}_1',{\bf k}_2',t_1,t_2,t_3)\equiv
\Theta(-t_1)\Theta(-t_2)\Theta(-t_3)
\Bigl\langle \tilde{\mathcal T}\Bigl[\hat c^{\dagger}_{{\bf k}_1'\uparrow}(t_1)
\hat c^{\dagger}_{1{\bf k}_2'\downarrow}(t_2)\Bigr] 
\hat c_{{\bf k}_2'\downarrow }(0)\hat c_{1{\bf k}_1'\uparrow}(t_3) \Bigr\rangle_{t_0}.
\label{eq_23}
\end{eqnarray}
The Fourier transforms of these correlation functions are defined by
\begin{equation}
G_{\sigma}({\bf k}_1,{\bf k}_2,\omega_1,\omega_2,\omega_3)=\int_{-\infty}^{\infty} dt_1\int_{-\infty}^{\infty} dt_2\int_{-\infty}^{\infty} dt_3
e^{i\omega_1t_1}e^{i\omega_2t_2}e^{-i\omega_3t_3}G_{\sigma}({\bf k}_1,{\bf k}_2,t_1,t_2,t_3).
\label{eq_j_1}
\end{equation}
Thus, the two-particle current is expressed in terms of these correlation functions as
\begin{eqnarray}
J({\bf k}_1,{\bf k}_2)&=&-\frac{1}{\hbar^4}|\gamma_{\uparrow}|^2|\gamma_{\downarrow}|^2
\Bigl[
iG_{\uparrow}({\bf k}_1-{\bf q}_{\uparrow},{\bf k}_2-{\bf q}_{\downarrow},\omega_{\uparrow}-\epsilon_{{\bf k}_1\uparrow}-i\eta,
\omega_{\downarrow}-\epsilon_{{\bf k}_2\downarrow}-i\eta,\omega_{\downarrow}-\epsilon_{{\bf k}_2\downarrow}+i\eta)\nonumber\\
&&~~~~~~~~~~+
iG_{\downarrow}({\bf k}_1-{\bf q}_{\uparrow},{\bf k}_2-{\bf q}_{\downarrow},\omega_{\downarrow}-\epsilon_{{\bf k}_1\uparrow}-i\eta,
\omega_{\uparrow}-\epsilon_{{\bf k}_2\downarrow}-i\eta,\omega_{\uparrow}-\epsilon_{{\bf k}_1\uparrow}+i\eta
)-{\rm c.c.}\Bigr]\nonumber \\
&=&
-\frac{2}{\hbar^4}|\gamma_{\uparrow}|^2|\gamma_{\downarrow}|^2
{\rm Im}\Bigl[
G_{\uparrow}({\bf k}_1-{\bf q}_{\uparrow},{\bf k}_2-{\bf q}_{\downarrow},\omega_{\uparrow}-\epsilon_{{\bf k}_1\uparrow},
\omega_{\downarrow}-\epsilon_{{\bf k}_2\downarrow})\nonumber\\
&&~~~~~~~~~~~~~~~~~~+
G_{\downarrow}({\bf k}_1-{\bf q}_{\uparrow},{\bf k}_2-{\bf q}_{\downarrow},\omega_{\uparrow}-\epsilon_{{\bf k}_1\uparrow},\omega_{\downarrow}-\epsilon_{{\bf k}_2\downarrow})\Bigr],
\end{eqnarray}
where we have denoted
\begin{eqnarray}
&&G_{\uparrow}({\bf k}_1-{\bf q}_{\uparrow},{\bf k}_2-{\bf q}_{\downarrow},\omega_{\uparrow}-\epsilon_{{\bf k}_1\uparrow},
\omega_{\downarrow}-\epsilon_{{\bf k}_2\downarrow}) \nonumber \\ 
&&~~\equiv  G_{\uparrow}({\bf k}_1-{\bf q}_{\uparrow},{\bf k}_2-{\bf q}_{\downarrow},\omega_{\uparrow}-\epsilon_{{\bf k}_1\uparrow}-i\eta,
\omega_{\downarrow}-\epsilon_{{\bf k}_2\downarrow}-i\eta,\omega_{\downarrow}-\epsilon_{{\bf k}_2\downarrow}+i\eta),
\\
&&G_{\downarrow}({\bf k}_1-{\bf q}_{\uparrow},{\bf k}_2-{\bf q}_{\downarrow},\omega_{\uparrow}-\epsilon_{{\bf k}_1\uparrow},\omega_{\downarrow}-\epsilon_{{\bf k}_2\downarrow}) \nonumber \\
&&~~\equiv
G_{\downarrow}({\bf k}_1-{\bf q}_{\uparrow},{\bf k}_2-{\bf q}_{\downarrow},\omega_{\uparrow}-\epsilon_{{\bf k}_1\uparrow}-i\eta,
\omega_{\downarrow}-\epsilon_{{\bf k}_2\downarrow}-i\eta,\omega_{\uparrow}-\epsilon_{{\bf k}_1\uparrow}+i\eta
).
\end{eqnarray}

\subsection{Lehmann representation}
In order to gain physical insight into the above results,
it is useful to white down the expression (\ref{eq_j_1}) for the two-particle current in the Lehman representation.
The statistical average $\langle \hat O \rangle_{t_0}$ of any operator can be expressed in terms of the energy eigenstates of $\hat H_1$ as
\begin{equation}
\langle \hat O \rangle_{t_0}=\sum_n \rho_n \langle n |\hat O |n\rangle,
\end{equation}
where $\rho_n$ is the diagonal element of the equilibrium statistical density operator in the energy representation.
In the grand-canonical ensemble, it is given as
\begin{equation}
\rho_n=\frac{\exp[-\beta(E_n-\mu N_n)]}{\Xi},~~~\Xi\equiv \sum_n \exp[-\beta(E_n-\mu N_n)].
\label{eq_29}
\end{equation}
In the energy representation, the matrix element of the Heisenberg operator is given by
\begin{equation}
\langle n |\hat O(t)|m\rangle=\langle n | e^{i\hat H t/\hbar}\hat O e^{-i \hat H t/\hbar}|m\rangle=e^{i(E_n-E_m)t/\hbar}\langle n |\hat O |m\rangle.
\end{equation}
Using this energy representation, we can express the two-particle correlation functions as
\begin{eqnarray}
&&G_{\uparrow}({\bf k}_1,{\bf k}_2,\omega_1,\omega_2)+G_{\downarrow}({\bf k}_1,{\bf k}_2,\omega_1,\omega_2)
=-\sum_n\sum_l\rho_n \frac{1}{\omega_1+\omega_2-(E_l-E_n)/\hbar-2i\eta}\nonumber \\
&&\times \left|\sum_m\frac{\bigl\langle l \bigl|\hat c_{{\bf k}_2'\downarrow}\bigr|m\bigr\rangle
\bigl\langle m \bigl| \hat c_{{\bf k}_1'\uparrow}\bigr|n\bigr\rangle}
{\omega_1-(E_m-E_n)/\hbar+i\eta}
-\sum_m\frac{\bigl\langle l\bigl|\hat c_{{\bf k}_1'\uparrow}\bigr|m\bigr\rangle
\bigl\langle m  \bigl| \hat c_{{\bf k}_2'\downarrow}\bigr|n\bigr\rangle
}{\omega_2-(E_m-E_n)/\hbar+i\eta}\right|^2.
\label{eq_G+}
\end{eqnarray}
We thus obtain the general expression for the two-particle current as
\begin{eqnarray}
J({\bf k}_1,{\bf k}_2)&=&
\frac{2\pi}{\hbar^4}|\gamma_{\uparrow}|^2|\gamma_{\downarrow}|^2
\sum_n\sum_l\rho_n \delta(\omega_1+\omega_2-(E_l-E_n)/\hbar)\nonumber \\
&&\times \left|\sum_m\frac{\bigl\langle l \bigl|\hat c_{{\bf k}_2'\downarrow}\bigr|m\bigr\rangle
\bigl\langle m \bigl| \hat c_{{\bf k}_1'\uparrow}\bigr|n\bigr\rangle}
{\omega_1-(E_m-E_n)/\hbar+i\eta}
-\sum_{m'}\frac{\bigl\langle l\bigl|\hat c_{{\bf k}_1'\uparrow}\bigr|m'\bigr\rangle
\bigl\langle m'  \bigl| \hat c_{{\bf k}_2'\downarrow}\bigr|n\bigr\rangle
}{\omega_2-(E_{m'}-E_n)/\hbar+i\eta}
\right|^2,
\label{eq_J_r}
\end{eqnarray}
where
\begin{equation}
{\omega_1=\omega_{\uparrow}-\epsilon_{{\bf k}_1\uparrow},
\omega_2=\omega_{\downarrow}-\epsilon_{{\bf k}_2\downarrow},
{\bf k}_1'={\bf k}_1-{\bf q}_{\uparrow},
{\bf k}_2'={\bf k}_2-{\bf q}_{\downarrow}}.
\end{equation}
The energy delta function in (\ref{eq_J_r}) arises from taking the limit $\eta\to 0$ in the common prefactor
of (\ref{eq_G+}).

The physical meaning of the energy eigenstates $|n\rangle, |m\rangle,|m'\rangle,|l\rangle$ are understood as follows:
$|n\rangle$ describes the initial state,
$|m\rangle$ describes the state where a ${\bf k}_1'\uparrow$ particle is subtracted from the initial state,
$|m'\rangle$ describes the state where a ${\bf k}_2'\downarrow$ particle is subtracted from the initial state, and 
$|l\rangle$ describes the state where  ${\bf k}_1'\uparrow$ particle  and ${\bf k}_2'\downarrow$ particle are
subtracted from the initial state.

\section{Two-particle spectrum in Fermi superfluid}
\subsection{BCS-Leggett's theory}
We now consider a uniform Fermi superfluid gas at $T=0$ to illustrate the most basic physics. 
In order to calculate the two-particle spectral function explicitly, we must specify a microscopic approximation. Here we use Leggett's theory \cite{T_Luxat_PRA2002} based on the mean-field treatment in the BCS-BEC crossover at $T=0$.
For this purpose, we work with the grand-canonical Hamiltonian defined by
\begin{equation}
\hat K_1=\hat H_1-\mu \hat N_1,
\end{equation}
where $\hat N_1\equiv \sum_{\sigma}\sum_{{\bf k}}\hat c_{{\bf k}\sigma}^{\dagger}\hat c_{{\bf k}\sigma}$ and $\mu$ is the chemical potential
of the system 1. 
It is convenient to introduce the Heisenberg operator defined in terms of the grand canonical Hamiltonian as
\begin{equation}
\hat c_{{\bf k}\sigma K}(t)\equiv e^{i\hat K_1 t/\hbar }\hat c_{{\bf k}\sigma} e^{-i\hat K_1 t/\hbar}
=e^{i\hat H_1 t/\hbar }e^{-i\mu\hat N_1 t/\hbar}\hat c_{{\bf k}\sigma} e^{i\mu\hat N_1 t/\hbar}
e^{-i\hat H_1 t/\hbar}.
\end{equation}
Here we have used the fact that the Hamiltonian commutes with the total number operator.
Using the identity $e^{-i\mu\hat N_1 t/\hbar}\hat c_{{\bf k}\sigma} e^{i\mu\hat N_1 t/\hbar}=e^{i\mu t/\hbar}\hat c_{{\bf k}\sigma}$, it is easily verified that
\begin{equation}
\hat c_{{\bf k}\sigma K}(t)=e^{i\mu t/\hbar}\hat c_{{\bf k}\sigma H}(t)~~~\mbox{or}~~~~
\hat c_{{\bf k}\sigma H}(t)=e^{-i\mu t/\hbar}\hat c_{{\bf k}\sigma K}(t).
\label{eq_38}
\end{equation}
The grand-canonical Hamiltonian for a uniform Fermi superfluid is given by
\begin{equation}
\hat K_1=\sum_{{\bf k},\sigma} \xi_{\bf k} \hat c_{{\bf k}\sigma}\hat c_{{\bf k}\sigma}
+U\sum_{{\bf k},{\bf k}',{\bf q}}\hat c^\dagger_{{\bf k}+{\bf q}/2\uparrow}\hat c^\dagger_{-{\bf k}'+{\bf q}/2\downarrow}\hat c_{-{\bf k}'+{\bf q}/2\downarrow}\hat c_{{\bf k}+{\bf q}/2\uparrow},
\end{equation}
where $\xi_{\bf k}=\epsilon_{\bf k}-\mu$ with $\epsilon_{\bf k}=\frac{\hbar {\bf k}^2}{2m}-\mu$. The pairing interaction $U=-\frac{4\pi \hbar^2 a_s }{m}$ is assumed to be tunable by a Feshbach resonance, which is related to the $s$-wave scattering length $a_s$.
We introduce the mean-field static superfluid order parameter $\Delta$ as,  
\begin{eqnarray}
  \Delta= U\sum_{\bf k}\langle \hat c_{{\bf k}\downarrow} \hat c_{{\bf k}\uparrow} \rangle. \label{delta_0}
   \end{eqnarray} 
   
As usual, we define the quasiparticle operators by the Bogoliubov transformation:
\begin{equation}
\begin{array}{l}
\hat c_{{\bf k}\uparrow}=u_{{\bf k}}\hat \alpha_{{\bf k}\uparrow}+v_{{\bf k}}\hat \alpha^{\dagger}_{-{\bf k}\downarrow}, \\
\hat c_{{\bf k}\downarrow}=u_{{\bf k}}\hat \alpha_{{\bf k}\downarrow}-v_{{\bf k}}\hat \alpha^{\dagger}_{-{\bf k}\uparrow}. \\
\end{array}
\end{equation}   
In terms of these quasiparticle operators, the grand canonical Hamiltonian can be written as
\begin{equation}
\hat K_1=\sum_{\bf k} E_{\bf k}(\hat\alpha^{\dagger}_{{\bf k}\uparrow}\hat\alpha_{{\bf k}\uparrow}+\hat\alpha^{\dagger}_{{\bf k}\downarrow}\hat\alpha_{{\bf k}\downarrow})
+\mbox{constant},
\end{equation}

where the quasiparticle energy is given by
\begin{equation}
E_{\bf k}=\sqrt{\left(\frac{\hbar^2 k^2}{2m}-\mu\right)^2+\Delta^2}.
\end{equation}
Denoting the Heisenberg operator defined by $\hat K_1$ as $\hat \alpha_{{\bf k}\sigma}(t)$, we have
\begin{equation}
\hat\alpha_{{\bf k}\sigma}(t)=\hat\alpha_{{\bf k}\sigma}e^{-iE_{\bf k} t/\hbar},~~~
\hat\alpha^{\dagger}_{{\bf k}\sigma}(t)=\hat\alpha_{{\bf k}\sigma}e^{iE_{\bf k} t/\hbar}.
\label{eq_41}
\end{equation}
In the usual (weak-coupling) BCS theory, the chemical potential $\mu$ can be taken to be equal to the Fermi energy $\epsilon_{\rm F}$. However, from the general point of view, $\mu$ should be determined by the equation for the number of fermions. Indeed, the chemical potential is found to remarkably deviate from the Fermi energy when the pairing interaction is strong. Within Leggett's theory, the gap function $\Delta$ and chemical potential $\mu$ are determined self-consistently from the following gap and number equations for a uniform Fermi gas. 
\begin{eqnarray}
1&=&-\frac{4\pi a_s}{m} \sum_{\bf k} \left[\frac{1}{2E_{\bf k}}-\frac{1}{\epsilon_k}\right],\label{gap_T=0}\\
N&=& \sum_{\bf k} \left[1-\frac{\xi_{\bf k}}{E_{\bf k}}\right]
.\label{N_T=0}
\end{eqnarray}

Assuming the BCS ground state that satisfies $\hat\alpha_{{\bf k}\sigma}|\Phi_0\rangle=0$, it is now straightforward to calculate the two particle
correlation functions (\ref{eq_22}) and (\ref{eq_23}). The final result for the two-particle current is given by
\begin{eqnarray}
&&J({\bf k}_1,{\bf k}_2,\omega^\prime_1,\omega^\prime_2)\nonumber \\
&&=\frac{2}{\hbar^4}|\gamma_{\uparrow}|^2|\gamma_{\downarrow}|^2
{\rm Im}\Biggl[
\frac{v_{{\bf k}_1'}^2 u_{{\bf k}'_1}^2}{\omega_1'+\omega_2'+2\mu/\hbar-2i\eta}\delta_{{\bf k}'_1,-{\bf k}'_2}
\nonumber \\
&&
+\frac{v_{{\bf k}_2'}^2 v_{{\bf k}'_1}^2}{\omega_1'+\omega_2'-(E_{k_1'}+E_{k_2'}-2\mu)/\hbar-2i\eta}
\Biggr]
\nonumber \\
&&~~~~~~~\times\left|
\frac{1}{\omega_1'-(E_{k_1'}-\mu)/\hbar-i\eta}
+\frac{1}{\omega_2'-(E_{k_2'}-\mu)/\hbar-i\eta}\right|^2,
\label{eq_f}
\end{eqnarray}
where
\begin{equation}
{\bf k}_1'={\bf k}_1-{\bf q}_{\uparrow},~~{\bf k}_2'={\bf k}_2-{\bf q}_{\downarrow},~~
\omega_1'=\omega_{\uparrow}-\epsilon_{{\bf k}_1}/\hbar,~~
\omega_2'=\omega_{\downarrow}-\epsilon_{{\bf k}_2}/\hbar.
\end{equation}
In Eq.~(\ref{eq_f}), the first term represents the contribution from the condensed pair components, while the second term is contribution from the uncorrelated pair states.

\subsection{The two-particle spectral function in the BCS-BEC crossover}
\begin{figure}[htbp]
\includegraphics[width=70mm]{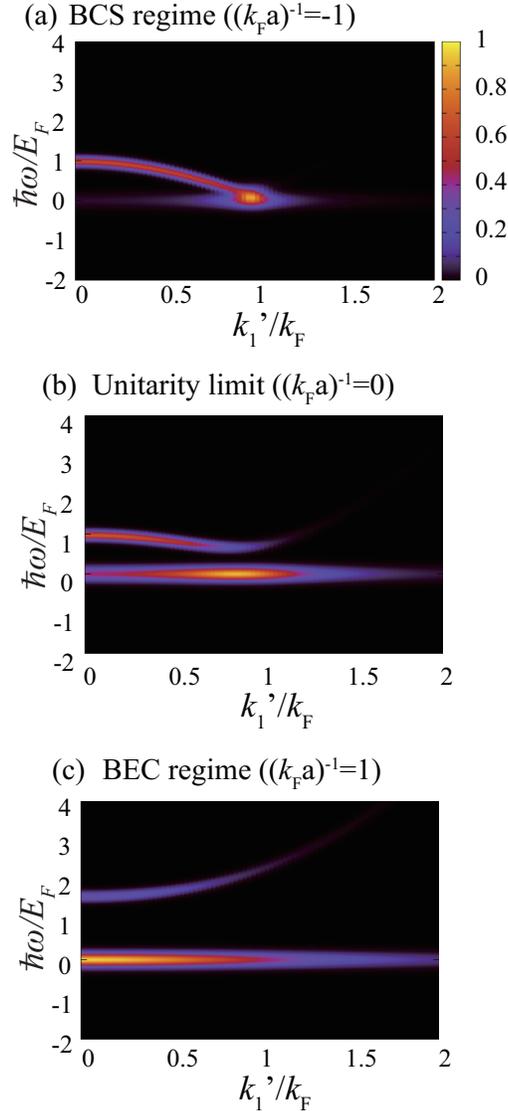}
\caption{\label{fig:one} (Color online) Intensity of two-particle current. The values of the pairing
interaction $(k_{\rm F} a_s)^{-1}$ are (a) -1, (b) 0, and (c) 1. }
\end{figure}

Hereafter we assume the case of $\omega_{\uparrow}=\omega_{\downarrow}=\omega$ for simplicity.
Figure~\ref{fig:one} shows intensity of the two-particle current $J({\bf k}_1,{\bf k}_2,\omega^\prime_1,\omega^\prime_2)$ for ${\bf k}_1^\prime=-{\bf k}_2^\prime$.
The peak at $\hbar \omega=0$ is the contribution of the condensed pair components, while the peak at $\hbar \omega\neq0$ is the contribution of uncorrelated pair states. The weight of condensed pair contribution increases with increasing pairing interaction.

\begin{figure}[htbp]
\includegraphics[width=50mm]{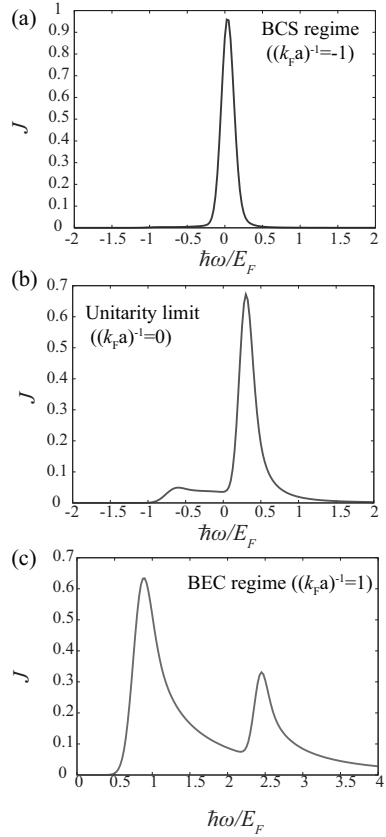}
\caption{\label{fig:sumJ} DPE current as a function of the energy $\omega$. The values of the pairing
interaction $(k_{\rm F} a_s)^{-1}$ are (a) -1, (b) 0, and (c) 1. }
\end{figure}

\begin{figure}[htbp]
\includegraphics[width=40mm]{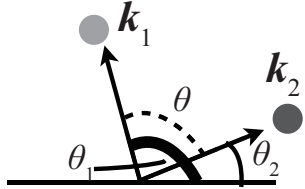}
\caption{\label{fig0} Sketch of  DPE process indicating two outgoing Fermions of with wave vector $\mathbf{k}_1$ and $\mathbf{k}_2$ as well as the emission angles $\theta_1$ and $\theta_2$.}
\end{figure} 

\begin{figure}[htbp]
\includegraphics[width=60mm]{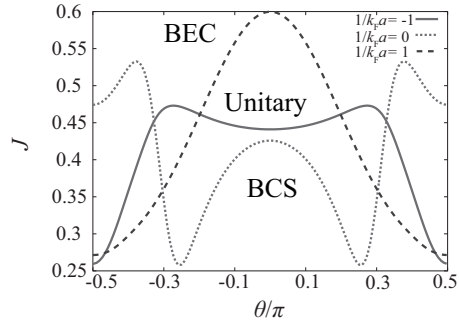}
\caption{\label{fig:sumtheta} The pair angular distributions as a
function of the relative scattering angle $\theta$. The values of the pairing
interaction $(k_{\rm F} a_s)^{-1}=$ -1, 0, and 1}
\end{figure}

Figure~\ref{fig:sumJ} shows the energy distribution of DPE current defined by $J(\omega)\equiv \sum_{{\bf k}_1,{\bf k}_2}J({\bf k}_1,{\bf k}_2,\omega)$.
We can see that the weight of condensed pair contribution increases with increasing pairing interaction. As we will see in Sec. IV, such a condensed pair contribution dose not appear in the single-particle spectrum even when considering the two-particle tunneling contribution. In contrast, DPE current as a function of the energy exhibits distinct  contributions of uncorrelated pair states and condensed pair components. However, in order to distinguish between weakly-bound Cooper pairs and tightly-bound molecules, it is not sufficient to see only the energy distributions of DPE current. From this point of view, we will show the angular distributions of DPE current.  

We define the angular distribution of DPE current as $J(\theta,\omega)=\sum_{{\bf k}_1,{\bf k}_2}J({\bf k}_1,{\bf k}_2,\omega)\delta(\theta_2-\theta_1-\theta)$, where $\theta_1$ and $\theta_2$ are the scattering angles of the emitted pair of fermions and $\theta$ is the relative scattering angle, as illustrated in Fig.~\ref{fig0}.

In Fig.~\ref{fig:sumtheta}, we plot the angular distribution of DPE current for $\omega=0$.
In the BCS side ($1/k_{\rm F}a_s=-1$), we see the double peak of DPE current, which corresponds to the case where the particles constructing a pair are respectively emitted. On the contrary, in the BEC side ($1/k_{\rm F}a_s=1$) the single peak appears, which means that the pairs emitted as molecules. This shows that one can distinguish between Cooper pairs and molecules from the angular distributions of DPE current.
We will show in the next section that such a condensed pair contribution cannot be seen by single-particle spectroscopy even when including the effect of the two-particle tunneling as the higher-order process in the tunneling Hamiltonian. 
\section{Effect of two particle tunneling in the Single-particle current}
In the case of superconductivity, effect of two-electron tunneling to single-particle current (i.e. Josephson current) was discussed by using the forth order perturbation theory. 
In this section, we give an analogous discussion of how the two-particle tunneling process affect the single-particle current in a superfluid Fermi gas in BCS-BEC crossover. 
In order to include the effect of two-particle tunneling, we have to calculate the forth order contribution. 
\subsection{General Formalism}
The single-particle current is defined as the rate of change of the single-particle density
\begin{eqnarray}
n_\sigma({\bf k},t)=\langle \hat b^\dagger_{{\bf k}\sigma}\hat b_{\bf k\sigma}  \rangle_t,
\end{eqnarray}
which is given as
\begin{eqnarray}
J_{\sigma}=\frac{d}{dt}n_{\sigma}({\bf k},t) = \frac{i}{\hbar} \langle[\hat b^\dagger_{\bf k\sigma}\hat b_{{\bf k}\sigma},\hat H ]\rangle_t.
\end{eqnarray}
The first order term gives the usual expression for the tunneling current, which is described in terms of the single-particle spectral function \cite{B_Mahan}.
Following the procedure similar to that in Sec.II, we obtain the higher-order contribution
\begin{eqnarray}
J_{\sigma}({\bf k},t)
&=&-\frac{1}{\hbar^4}\sum_{\sigma^\prime}\sum_{k^\prime}|\gamma_\sigma|^2|\gamma_{\sigma^\prime}|^2
G_{\sigma\sigma^\prime}(k^\prime,k_\sigma,\omega_{\sigma^{\prime}}^{\prime\prime}-i\eta,\omega_\sigma^\prime-i\eta,\omega_{\sigma^{\prime}}^{\prime\prime}+i\eta)
\nonumber \\
&&
-\frac{1}{\hbar^4}\sum_{\sigma^\prime}\sum_{k^\prime}|\gamma_\sigma|^2|\gamma_{\sigma^\prime}|^2
G'_{\sigma\sigma^\prime}(k^\prime,k_\sigma,\omega_{\sigma^{\prime}}^{\prime\prime}-i\eta,\omega_\sigma^\prime-i\eta,\omega_{\sigma^{\prime}}^{\prime\prime}+i\eta)
\nonumber \\
&&
-\frac{1}{\hbar^4}\sum_{\sigma^\prime}\sum_{k^\prime}|\gamma_\sigma|^2|\gamma_{\sigma^\prime}|^2
G''_{\sigma\sigma^\prime}(k^\prime,k_\sigma,\omega_{\sigma^{\prime}}^{\prime\prime}-i\eta,\omega_\sigma^\prime-i\eta,\omega_{\sigma^{\prime}}^{\prime\prime}+i\eta),
\end{eqnarray}
where we introduced the three kinds of the correlation functions: 
\begin{eqnarray}
iG_{\sigma\sigma^\prime}(k^\prime,k_\sigma,t_1,t_2,t_3)
&=&      \langle \tilde{\mathcal T} \left[
  \hat c^\dagger_{{\bf k}^{\prime}\sigma^{\prime}}(t_1) 
\hat c^\dagger_{{\bf k}-{\bf q}_{\sigma}\sigma}(t_2)
\right]
\hat c_{{\bf k}-{\bf q}_{\sigma}\sigma}(0)
\hat c_{{\bf k}^{\prime}\sigma^{\prime}}(t_3)
\rangle_{t_0}\theta (-t_1)\theta (-t_2)\theta (-t_3),
 \nonumber \\&&
\\
 iG'_{\sigma\sigma^\prime}(k^\prime,k_\sigma,t_1,t_2,t_3)
 &=&-\langle
\hat c^\dagger_{{\bf k}_{\sigma}\sigma}(t_2)
\hat c_{{\bf k}_{\sigma}\sigma}(0)
\hat c^\dagger_{{\bf k}^{\prime}\sigma^{\prime}}(t_1) 
\hat c_{{\bf k}^{\prime}\sigma^{\prime}}(t_3)
\rangle_{t_0}
\theta (t_1-t_3)  \theta (-t_1)\theta (-t_2)\theta (-t_3),
 \nonumber \\&&
\\
iG''_{\sigma\sigma^\prime}(k^\prime,k_\sigma,t_1,t_2,t_3)
   &=&
  \theta(-t_1)\theta(-t_2)\theta(-t_3) \theta (t_3-t_1)
 \nonumber \\&&
    \times\Bigl\{
-   \theta (t_1-t_2)\langle
\hat c^\dagger_{{\bf k}-{\bf q}_{\sigma}\sigma}(t_2)
\hat c^\dagger_{{\bf k}^{\prime}\sigma^{\prime}}(t_1)
\hat c_{{\bf k}^{\prime}\sigma^{\prime}}(t_3)
\hat c_{{\bf k}-{\bf q}_{\sigma}\sigma}(0)\rangle_{t_0}
 \nonumber \\&&
+   \theta (t_3-t_2)   \theta (t_2-t_1)\langle
\hat c^\dagger_{{\bf k}^{\prime}\sigma^{\prime}}(t_1)
\hat c^\dagger_{{\bf k}-{\bf q}_{\sigma}\sigma}(t_2)
\hat c_{{\bf k}^{\prime}\sigma^{\prime}}(t_3)
\hat c_{{\bf k}-{\bf q}_{\sigma}\sigma}(0)\rangle_{t_0}
 \nonumber \\&&
-   \theta (t_2-t_3)\langle
\hat c^\dagger_{{\bf k}^{\prime}\sigma^{\prime}}(t_1)
\hat c_{{\bf k}^{\prime}\sigma^{\prime}}(t_3)
\hat c^\dagger_{{\bf k}-{\bf q}_{\sigma}\sigma}(t_2)
\hat c_{{\bf k}-{\bf q}_{\sigma}\sigma}(0)\rangle_{t_0}
  \Bigl\}.
\end{eqnarray} 

Using the energy representation described in Sec.~III, one can also express the two-particle contribution to the single-particle current function in the Lehmann representation
\begin{eqnarray}
&&J_{\sigma}(\omega)
=\frac{1}{\hbar^4}\sum_{\sigma^\prime}\sum_{k^\prime}|\gamma_\sigma|^2|\gamma_{\sigma^\prime}|^2\sum_{n,m,l,k} \rho_n 
\nonumber \\ &&
\Bigg[
-\frac{\langle n |
\hat c^\dagger_{{\bf k}-{\bf q}_{\sigma}\sigma}|m\rangle 
\langle m |\hat c_{{\bf k}-{\bf q}_{\sigma}\sigma}|l\rangle 
 }{\omega_{1\sigma'}-(E_k-E_n)/\hbar-i\eta}
\frac{\langle l |\hat c^\dagger_{{\bf k}^{\prime}\sigma^{\prime}} |k\rangle 
\langle k |\hat c_{{\bf k}^{\prime}\sigma^{\prime}}
|n\rangle}{\omega_{2\sigma}+(E_n-E_m)/\hbar-i\eta}
\frac{1}{(E_l+E_n-2E_k)/\hbar-2i\eta}
\nonumber \\ &&
+\frac{1}{(\omega_{1\sigma'}+\omega_{2\sigma}+(E_n-E_l)/\hbar-2i\eta)}
\frac{\langle n | \hat c^\dagger_{{\bf k}^{\prime}\sigma^{\prime}} |m \rangle
\langle m | \hat c^\dagger_{{\bf k}-{\bf q}_{\sigma}\sigma}|l\rangle  }
{\omega_{1\sigma'}+(E_m-E_l)/\hbar-i\eta}
\frac{\langle l | \hat c_{{\bf k}-{\bf q}_{\sigma}\sigma} |k \rangle 
\langle k | \hat c_{{\bf k}^{\prime}\sigma^{\prime}}|n \rangle }
{\omega_{1\sigma'}+(E_k-E_n)/\hbar+i\eta}
\nonumber \\ &&
+\frac{1}{(\omega_{1\sigma'}+\omega_{2\sigma}+(E_n-E_l)/\hbar-2i\eta)}
\frac{\langle n | \hat c^\dagger_{{\bf k}-{\bf q}_{\sigma}\sigma} |m \rangle 
\langle m |\hat c^\dagger_{{\bf k}^{\prime}\sigma^{\prime}} |l\rangle }
{\omega_{2\sigma}+(E_m-E_l)/\hbar-i\eta}
\frac{\langle l | \hat c_{{\bf k}-{\bf q}_{\sigma}\sigma} |k \rangle 
\langle k | \hat c_{{\bf k}^{\prime}\sigma^{\prime}}|n \rangle}
{\omega_{1\sigma'}+(E_k-E_n)/\hbar+i\eta}
\nonumber \\ &&
    +\frac{\langle n |\hat c^\dagger_{{\bf k}-{\bf q}_{\sigma}\sigma}|m\rangle
\langle m |\hat c^\dagger_{{\bf k}^{\prime}\sigma^{\prime}}|l\rangle
}{(\omega_{1\sigma'}- (E_l-E_k)/\hbar-i\eta)}\frac{\langle l |\hat c_{{\bf k}^{\prime}\sigma^{\prime}}|k\rangle
\langle k |\hat c_{{\bf k}-{\bf q}_{\sigma}\sigma}|n\rangle}{\omega_{2\sigma}+(E_n-E_k)/\hbar-i\eta}
\frac{1}{(E_k-E_m)/\hbar}
  \Bigg].
\end{eqnarray}

Using BCS-Leggett's theory as in Sec. III A, we obtain  
\begin{eqnarray}
&&J_\uparrow=
\frac{2}{\hbar^4}|\gamma_{\uparrow}|^2|\gamma_{\uparrow}|^2
\nonumber \\ &&~~~~~~\times
{\rm Im}
\Biggl[
\frac{v^2_{{\bf k}_\uparrow} u^2_{{\bf k}_\uparrow}}{\omega_{1\uparrow}+\omega_{2\uparrow}+2\mu/\hbar-2i\eta}\delta_{{\bf k}_\uparrow,{\bf k}'}
+\frac{v_{{\bf k}_\uparrow}^2 v_{{\bf k}'}^2}{\omega_{1\uparrow}+\omega_{2\uparrow}-(E_{k^\prime}+E_{k_\uparrow}-2\mu)/\hbar-2i\eta}
\Biggr]
\nonumber \\
&&~~~~~~~~~~~~~~~~~~~~~~~~~\times\left|
\frac{1}{\omega_{1\uparrow}-(E_{k^\prime}-\mu)/\hbar-i\eta}
+\frac{1}{\omega_{2\uparrow}-(E_{k_\uparrow}-\mu)/\hbar-i\eta}\right|^2
\nonumber \\ &&
+\frac{2}{\hbar^4}|\gamma_{\uparrow}|^2|\gamma_{\downarrow}|^2
\nonumber \\ &&~~~~~~\times
{\rm Im}\Biggl[
\frac{v^2_{{\bf k}_\uparrow} u^2_{{\bf k}_\uparrow}}{(\omega_{1\downarrow}+\omega_{2\uparrow}+2\mu/\hbar-2i\eta)}\delta_{{\bf k}_\uparrow,{\bf k}'}
+\frac{v_{{\bf k}_\uparrow}^2 v_{{\bf k}'}^2}{\omega_{1\downarrow}+\omega_{2\uparrow}-(E_{k^\prime}+E_{k_\uparrow}-2\mu)/\hbar-2i\eta}
\Biggr]
\nonumber \\
&&~~~~~~~~~~~~~~~~~~~~~~~~~\times\left|
\frac{1}{\omega_{1\downarrow}-(E_{k^\prime}-\mu)/\hbar-i\eta}
+\frac{1}{\omega_{2\uparrow}-(E_{k_\uparrow}-\mu)/\hbar-i\eta}\right|^2.
\label{Eq_s}
\end{eqnarray}
One can see that the first terms in the square brackets represent the contribution from the tunneling a condensate pair, while the second terms represent the contribution of uncorrelated pair states.
\subsection{numerical calculations}
\begin{figure}[htbp]
\includegraphics[width=70mm]{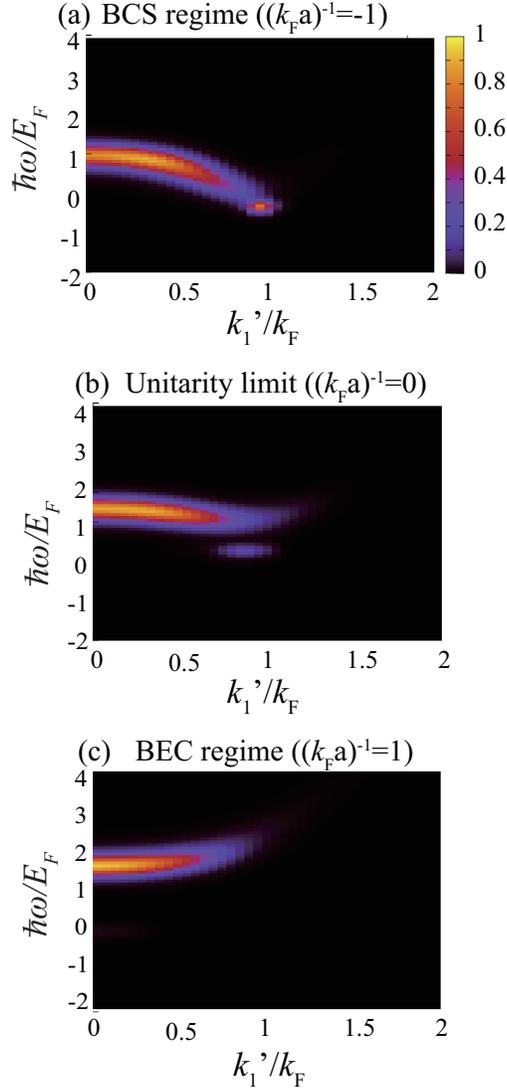}
\caption{\label{fig:single} (Color online) Intensity of single-particle current. The values of the pairing
interaction $(k_{\rm F} a_s)^{-1}$ are (a) -1, (b) 0, and (c) 1. }
\end{figure}

Figure~\ref{fig:single} shows intensity of the two-particle contribution to the single-particle current in the BCS-BCS crossover in Fig.~\ref{fig:single}. In the BEC regime, the contribution of uncorrelated pair states is much lager than that of condensed pairs. 
In the BCS side, however, we can see a small peak from contribution of condensed pairs near $k/k_F=1$. The weight of condensed pair contribution decreases with increasing pairing interaction. We note that this behavior of condensed pair contribution is very deferent from that of condensed pair contribution of DPE current, where the weight of condensed pair contribution increases with increasing pairing interaction. 
In contrast to the two-particle current, one can only see the two-particle
contribution to the single-particle current in the unitarity regime.

\section{Conclusion}
We provided a general formalism for DPE current from superfluid Fermi gases within the framework of the time-dependent perturbation theory. Using this formalism,  
we studied DPE current in superfluid Fermi gases in the BCS-BEC crossover at $T=0$
within the framework of BCS-Leggett's theory. From the intensity of two-particle spectral densities and energy distributions of DPE current, 
we can identify the contribution of condensed pairs and uncorrelated states with the energy of peaks.  
The peak at $\hbar \omega=0$ is the contribution of condensed pairs and other peaks are contribution of uncorrelated states.
DPE current as a function of the energy also showed the very deferent  contribution of uncorrelated pair states and condensed pair components. 

We also calculated the angular distributions of DPE current in the BCS-BEC crossover. In the BCS side ($1/k_{\rm F}a_s=-1$), we show the double peak of DPE current, which corresponds the particles constructing a pair respectively emitted. While in the BEC side ($1/k_{\rm F}a_s=1$) we can see the single peak, which means the pairs emitted as molecules. 

For comparison, we showed the contribution of the two-particle tunneling process to the single-particle current. We found that the contribution of uncorrelated pair states is always much lager than that of condensed pairs. A small peak from contribution of condensed pairs appears near $k/k_F=1$ in the BCS side. 

In summary, the present study showed the possibility of distinguishing between weakly-bound Cooper pairs and tightly-bound molecules. 
We hope that these results will simulate in further experiment.

\begin{acknowledgments}
We thank S. Tsuchiya for valuable comments. This work was supported by a Grant-in-Aid from JSPS.
\end{acknowledgments}
\appendix
\section{General formulation of the two-particle current}
We derive the third-order expressions (\ref{eq45a}) and (\ref{eq45b})
for the correlation functions $\langle \hat F_{\sigma}\rangle $.
According to (\ref{eq18}), we need the following expectation value:
\begin{eqnarray}
&&\left\langle {\mathcal T}\left[
\hat V(t')\hat V(t'')\hat V(t''')\hat F({\bf k}_1,{\bf k}_2,{\bf k}_3,{\bf k}_4)(t)\right]\right\rangle_{t_0}
=\sum_{\sigma''}\sum_{{\bf k}''}\sum_{\sigma'}\sum_{{\bf k}'}\sum_{\sigma}\sum_{{\bf k}}e^{\eta t'''}e^{\eta t''}e^{\eta t'}
\nonumber \\
&&\times \Bigl\langle {\mathcal T}\Bigl[\left(\gamma_{\sigma}e^{-i \omega_{\sigma}t'}
\hat b^{\dagger}_{{\bf k}+{\bf q}_{\sigma}\sigma}(t')
\hat c_{{\bf k}\sigma}(t')+{\rm H.c.}\right)
\left(\gamma_{\sigma'}e^{-i \omega_{\sigma'}t''}
\hat b^{\dagger}_{{\bf k}'+{\bf q}_{\sigma'}\sigma'}(t'')
\hat c_{{\bf k}'\sigma'}(t'')+{\rm H.c.}\right)\nonumber \\
&&\times
\left(\gamma_{\sigma''}e^{-i \omega_{\sigma''}t'''}
\hat b^{\dagger}_{{\bf k}''+{\bf q}_{\sigma''}\sigma''}(t''')
\hat c_{{\bf k}''\sigma''}(t''')+{\rm H.c.}\right)
\hat b^{\dagger}_{ {\bf k}_1\uparrow}(t)
\hat b^{\dagger}_{{\bf k}_2\downarrow }(t)
\hat b_{{\bf k}_3\downarrow }(t)
\hat c_{{\bf k}_4\uparrow }(t)
\Bigr]
\Bigr\rangle_{t_0}.
\label{eq:a1}
\end{eqnarray}
Collecting the terms that makes non-vanishing contributions, we obtain
\begin{eqnarray}
&&\left\langle {\mathcal T}\left[
\hat V(t')\hat V(t'')\hat V(t''')\hat F({\bf k}_1,{\bf k}_2,{\bf k}_3,{\bf k}_4)(t)\right]\right\rangle_{t_0}
\nonumber \\
&&=3\sum_{\sigma''}\sum_{{\bf k}''}\sum_{\sigma'}\sum_{{\bf k}'}\sum_{\sigma}\sum_{{\bf k}}e^{\eta t'''}e^{\eta t''}e^{\eta t'}
\gamma_{\sigma}^*\gamma_{\sigma'}^*\gamma_{\sigma''}
e^{i \omega_{\sigma}t'}e^{i \omega_{\sigma'}t''}e^{-i \omega_{\sigma''}t'''}
\nonumber \\
&&\times \Bigl\langle {\mathcal T}\Bigl[
\hat c^{\dagger}_{{\bf k}\sigma}(t')
\hat b_{{\bf k}+{\bf q}_{\sigma}\sigma}(t')
\hat c^{\dagger}_{{\bf k}'\sigma'}(t'')
\hat b_{{\bf k}'+{\bf q}_{\sigma'}\sigma'}(t'')
\hat b^{\dagger}_{{\bf k}''+{\bf q}_{\sigma''}\sigma''}(t''')
\hat c_{{\bf k}''\sigma''}(t''')
\nonumber \\
&&~~~~~~~~~~~~~\times
\hat b^{\dagger}_{ {\bf k}_1\uparrow}(t)
\hat b^{\dagger}_{{\bf k}_2\downarrow }(t)
\hat b_{{\bf k}_3\downarrow }(t)
\hat c_{{\bf k}_4\uparrow }(t)
\Bigr]\Bigr\rangle_{t_0}.
\nonumber \\
&&
\end{eqnarray}
Under the assumption that two systems are initially uncoupled, we obtain
\begin{eqnarray}
&&\left\langle {\mathcal T}\left[
\hat V(t')\hat V(t'')\hat V(t''')\hat F({\bf k}_1,{\bf k}_2,{\bf k}_3,{\bf k}_4)(t)\right]\right\rangle_{t_0}
\nonumber \\
&&=3\sum_{\sigma''}\sum_{{\bf k}''}\sum_{\sigma'}\sum_{{\bf k}'}\sum_{\sigma}\sum_{{\bf k}}e^{\eta t'''}e^{\eta t''}e^{\eta t'}
\gamma_{\sigma}^*\gamma_{\sigma'}^*\gamma_{\sigma''}
e^{i \omega_{\sigma}t'}e^{i \omega_{\sigma'}t''}e^{-i \omega_{\sigma''}t'''}
\nonumber \\
&&\times \Bigl\langle {\mathcal T}\Bigl[
\hat b_{{\bf k}+{\bf q}_{\sigma}\sigma}(t')
\hat b_{{\bf k}'+{\bf q}_{\sigma'}\sigma'}(t'')
\hat b^{\dagger}_{{\bf k}''+{\bf q}_{\sigma''}\sigma''}(t''')
\hat b^{\dagger}_{ {\bf k}_1\uparrow}(t)
\hat b^{\dagger}_{{\bf k}_2\downarrow }(t)
\hat b_{{\bf k}_3\downarrow }(t)\Bigr]\Bigr\rangle_{t_0}\nonumber \\
&& \times\Bigl\langle {\mathcal T}\Bigl[\hat c^{\dagger}_{{\bf k}\sigma}(t')\hat c^{\dagger}_{{\bf k}'\sigma'}(t'')
\hat c_{{\bf k}''\sigma''}(t''')\hat c_{{\bf k}_4\uparrow }(t)
\Bigr]\Bigr\rangle_{t_0}.
\label{eq:a3}
\end{eqnarray}

Let us now assume that $\hat H_2$ takes the non-interacting form:
\begin{equation}
\hat H_2=\sum_{\sigma}\sum_{{\bf k}}\epsilon_{{\bf k}\sigma}b^{\dagger}_{{\bf k}\sigma}\hat b_{{\bf k}\sigma}.
\end{equation}
In this case, we can use the Wick's theorem for the state 2,  and use
\begin{equation}
\hat b_{{\bf k}\sigma}(t)=\hat b_{{\bf k}\sigma}e^{-i\epsilon_{{\bf k}\sigma}t/\hbar},~~
\hat b^{\dagger}_{{\bf k}\sigma}(t)=\hat b_{{\bf k}\sigma}e^{i\epsilon_{{\bf k}\sigma}t/\hbar},
\end{equation}
Furthermore, we assume that there is no particles in the initial state at $t=t_0$. and thus use 
\begin{equation}
\langle \hat b_{{\bf k}\sigma}\hat b^{\dagger}_{{\bf k}'\sigma'}\rangle_{t_0}=\delta_{\sigma\sigma'}
\delta_{{\bf k}{\bf k}'}.
\end{equation}
With these assumptions, we obtain 
\begin{eqnarray}
&&\Bigl\langle {\mathcal T}\Bigl[
\hat b_{{\bf k}+{\bf q}_{\sigma}\sigma}(t')\hat b_{{\bf k}'+{\bf q}_{\sigma'}\sigma'}(t'')
\hat b^{\dagger}_{{\bf k}''+{\bf q}_{\sigma''}\sigma''}(t''')\hat b^{\dagger}_{ {\bf k}_1\uparrow}(t)
\hat b^{\dagger}_{{\bf k}_2\downarrow }(t)\hat b_{{\bf k}_3\downarrow }(t)\Bigr]\Bigr\rangle_{t_0}
 \nonumber \\
&&=
\Biggl
(\Theta(t',t)\Theta(t'',t)
\delta_{\sigma\uparrow}\delta_{{\bf k},{\bf k}_1-{\bf q}_{\uparrow}}
e^{-i\epsilon_{{\bf k}_1\uparrow}(t'-t)/\hbar}
\delta_{\sigma'\downarrow}\delta_{{\bf k}',{\bf k}_2-{\bf q}_{\downarrow}}
e^{-i\epsilon_{{\bf k}_2\downarrow}(t''-t)/\hbar}
 \nonumber \\
&&~~~~
-\Theta(t',t)\Theta(t'',t)
\delta_{\sigma\downarrow}\delta_{{\bf k},{\bf k}_2-{\bf q}_{\downarrow}}
e^{-i\epsilon_{{\bf k}_2\downarrow}(t'-t)/\hbar}
\delta_{\sigma'\uparrow}\delta_{{\bf k}',{\bf k}_1-{\bf q}_{\uparrow}}
e^{-i\epsilon_{{\bf k}_1\uparrow}(t''-t)/\hbar}\Biggr)
\nonumber \\
&&~~~~\times
\Theta(t,t''') 
\delta_{\sigma''\downarrow}\delta_{{\bf k}'',{\bf k}_3-{\bf q}_{\downarrow}}
e^{-i\epsilon_{{\bf k}_3\downarrow}(t-t''')/\hbar}.
\end{eqnarray}
Using this result in (\ref{eq:a3}), we obtain
\begin{eqnarray}
&&\left\langle {\mathcal T}\left[
\hat V(t')\hat V(t'')\hat V(t''')\hat F({\bf k}_1,{\bf k}_2,{\bf k}_3,{\bf k}_4)(t)\right]\right\rangle_{t_0}
\nonumber \\
&&=
3\gamma_{\uparrow}^*|\gamma_{\downarrow}|^2
e^{\eta t'''}e^{\eta t''}e^{\eta t'}
e^{-i \omega_{\downarrow}t'''}
\nonumber \\
&&\times\Biggl
(\Theta(t',t)\Theta(t'',t)\Theta(t,t''') 
e^{i \omega_{\uparrow}t'}e^{i \omega_{\downarrow}t''}
e^{-i\epsilon_{{\bf k}_1\uparrow}(t'-t)/\hbar}
e^{-i\epsilon_{{\bf k}_2\downarrow}(t''-t)/\hbar}e^{-i\epsilon_{{\bf k}_3\downarrow}(t-t''')/\hbar}
 \nonumber \\
&&~~~~\times
\Bigl\langle {\mathcal T}\Bigl[\hat c^{\dagger}_{{\bf k}_1-{\bf q}_{\uparrow}\uparrow}(t')\hat c^{\dagger}_{{\bf k}_2-{\bf q}_{\downarrow}\downarrow}(t'')
\hat c_{{\bf k}_3-{\bf q}_{\downarrow}\downarrow}(t''')\hat c_{{\bf k}_4\uparrow }(t)
\Bigr]\Bigr\rangle_{t_0} \nonumber \\
&&~~~~
+\Theta(t',t)\Theta(t'',t)\Theta(t,t''') 
e^{i \omega_{\downarrow}t'}e^{i \omega_{\uparrow}t''}
e^{-i\epsilon_{{\bf k}_2\downarrow}(t'-t)/\hbar}
e^{-i\epsilon_{{\bf k}_1\uparrow}(t''-t)/\hbar}
e^{-i\epsilon_{{\bf k}_3\downarrow}(t-t''')/\hbar}\nonumber \\
&&~~~~\times
\Bigl\langle {\mathcal T}\Bigl[\hat c^{\dagger}_{{\bf k}_1-{\bf q}_{\uparrow}\uparrow}(t'')
\hat c^{\dagger}_{{\bf k}_2-{\bf q}_{\downarrow}\downarrow}(t')
\hat c_{{\bf k}_3-{\bf q}_{\downarrow}\downarrow}(t''')\hat c_{{\bf k}_4\uparrow }(t)
\Bigr]\Bigr\rangle_{t_0}\Biggr).
\end{eqnarray}

Using the above result in Eq.~(\ref{eq:a1}), we obtain
\begin{eqnarray}
&&\langle \hat F_{\uparrow}({\bf k}_1,{\bf k}_2,{\bf k}_3,{\bf k}_4) \rangle_t^{(3)}
\nonumber \\&&
=
\left(-\frac{i}{\hbar}\right)^3
\gamma_{\uparrow}^*|\gamma_{\downarrow}|^2
e^{i \omega_{\uparrow}t}
\oint_{t_0}^{t_0} dt' \oint_{t_0}^{t_0} dt'' \oint_{t_0}^{t_0} dt''' 
e^{\eta t'''}e^{\eta t''}e^{\eta t'}\Theta(t',t)\Theta(t'',t)\Theta(t,t''') 
\nonumber \\
&&\times
e^{i(\omega_{\uparrow}-\epsilon_{{\bf k}_1\uparrow}/\hbar)(t'-t)}
e^{i(\omega_{\downarrow}-\epsilon_{{\bf k}_2\downarrow}/\hbar)(t''-t)}
e^{-i(\omega_{\downarrow}-\epsilon_{{\bf k}_3\downarrow}/\hbar)(t'''-t)}
 \nonumber \\
&&~~~~\times
\Bigl\langle {\mathcal T}\Bigl[\hat c^{\dagger}_{{\bf k}_1-{\bf q}_{\uparrow}\uparrow}(t')\hat c^{\dagger}_{{\bf k}_2-{\bf q}_{\downarrow}\downarrow}(t'')
\hat c_{{\bf k}_3-{\bf q}_{\downarrow}\downarrow}(t''')\hat c_{{\bf k}_4\uparrow }(t)
\Bigr]\Bigr\rangle_{t_0}.
\label{eq8e}
\end{eqnarray}
Here we recall that the time integral goes from $t_0$ to $t$ on the chronological branch and goes back from $t$
to $t_0$ on the antichronological branch, and thus one always have $t',t'',t''<t$. On the other hand, in order
to make a non-vanishing contribution one must have $\Theta(t',t)\Theta(t'',t)\Theta(t,t''') =1$. Therefore $t',t''$ should be on the anti-chronological branch and $t'''$ should be on the chronological branch.
We can thus project on the contour-path integral to the real time axis as
\begin{eqnarray}
&&\langle \hat F_{\uparrow}({\bf k}_1,{\bf k}_2,{\bf k}_3,{\bf k}_4) \rangle_t^{(3)}
\nonumber \\ &&=
\left(-\frac{i}{\hbar}\right)^3
\gamma_{\uparrow}^*|\gamma_{\downarrow}|^2
e^{i \omega_{\uparrow}t}
\int_{t_0}^{t} dt' \int_{t_0}^{t} dt'' \int_{t_0}^{t} dt''' 
e^{\eta t'''}e^{\eta t''}e^{\eta t'}
\nonumber \\
&&\times
e^{i(\omega_{\uparrow}-\epsilon_{{\bf k}_1\uparrow}/\hbar)(t'-t)}
e^{i(\omega_{\downarrow}-\epsilon_{{\bf k}_2\downarrow}/\hbar)(t''-t)}
e^{-i(\omega_{\downarrow}-\epsilon_{{\bf k}_3\downarrow}/\hbar)(t'''-t)}
 \nonumber \\
&&~~~~\times
\Bigl\langle \tilde{\mathcal T}\Bigl[\hat c^{\dagger}_{{\bf k}_2-{\bf q}_{\downarrow}\downarrow}(t'')
\hat c^{\dagger}_{{\bf k}_1-{\bf q}_{\uparrow}\uparrow}(t')\Bigr] 
\hat c_{{\bf k}_4\uparrow }(t)\hat c_{{\bf k}_3-{\bf q}_{\downarrow}\downarrow}(t''') \Bigr\rangle_{t_0},
\label{eq8f}
\end{eqnarray}
Introducing the relative time coordinates
\begin{equation}
t_1=t'-t,~~t_2=t''-t,~~t_3=t'''-t,
\end{equation}
we obtain
\begin{eqnarray}
&&\langle \hat F_{\uparrow}({\bf k}_1,{\bf k}_2,{\bf k}_2,{\bf k}_1-{\bf q}_{\uparrow}) \rangle_t^{(3)}=
\left(-\frac{i}{\hbar}\right)^3
\gamma_{\uparrow}^*|\gamma_{\downarrow}|^2
e^{i \omega_{\uparrow}t}
\int_{t_0}^{0} dt_1 \int_{t_0}^{0} dt_2 \int_{t_0}^{0} dt_3
\nonumber \\
&&\times e^{\eta(t_1+t)}e^{\eta(t_2+t)}e^{\eta(t_3+t)}
e^{i(\omega_{\uparrow}-\epsilon_{{\bf k}_1\uparrow}/\hbar)t_1}
e^{i(\omega_{\downarrow}-\epsilon_{{\bf k}_2\downarrow}/\hbar)(t_2-t_3)}
 \nonumber \\
&&~~~~\times
\Bigl\langle \tilde{\mathcal T}\Bigl[\hat c^{\dagger}_{{\bf k}_2-{\bf q}_{\downarrow}\downarrow}(t_2)
\hat c^{\dagger}_{{\bf k}_1-{\bf q}_{\uparrow}\uparrow}(t_1)\Bigr] 
\hat c_{{\bf k}_1-{\bf q}_{\uparrow}\uparrow }(0)\hat c_{{\bf k}_2-{\bf q}_{\downarrow}\downarrow}(t_3) \Bigr\rangle_{t_0}.
\label{eq45}
\end{eqnarray}
Here we have made use of the fact that in equilibrium $\langle \hat A(t_1)\hat B(t_2)\rangle_{t_0}=\langle \hat A(t_1-t_2)\hat B(0)\rangle_{t_0}$
and so on.

\section{Lehmann representation of the two-particle current}
Using the energy representation (\ref{eq_29}), the two-particle correlation functions can be expressed as
\begin{eqnarray}
&&iG_{\uparrow}({\bf k}_1',{\bf k}_2',t_1,t_2,t_3)
\nonumber \\
&&=\Theta(-t_1)\Theta(-t_2)\Theta(-t_3)\sum_n\sum_m\sum_l\sum_k \rho_n \nonumber \\
&&~~\times \Biggl[\Theta(t_1-t_2)e^{i(E_n-E_m)t_2/\hbar} e^{i(E_m-E_l)t_1/\hbar}e^{i(E_k-E_n)t_3/\hbar}  
\nonumber \\&&~~\times 
\bigl\langle n \bigl|\hat c^{\dagger}_{{\bf k}_2'\downarrow}\bigr |m \bigr\rangle 
\bigl\langle m \bigl |  \hat c^{\dagger}_{{\bf k}_1'\uparrow}\bigr|l\bigr\rangle
\bigl\langle l \bigl |\hat c_{{\bf k}_1'\uparrow }\bigr|k \bigr\rangle \bigl\langle k \bigl | \hat c_{{\bf k}_2'\downarrow}
\bigr|n \bigr\rangle
\nonumber \\
&&~~~~~~-\Theta(t_2-t_1)
e^{i(E_n-E_m)t_1/\hbar} e^{i(E_m-E_l)t_2/\hbar}e^{i(E_k-E_n)t_3/\hbar} 
\nonumber \\&&~~\times  
\bigl\langle n \bigl|\hat c^{\dagger}_{{\bf k}_1'\uparrow}\bigr|m\bigr\rangle
\bigl\langle m\bigl |\hat c^{\dagger}_{{\bf k}'_2\downarrow}\bigr|l \bigr\rangle \bigl\langle l \bigl |
\hat c_{{\bf k}'_1\uparrow } \bigr | k \bigr\rangle\bigl\langle k \bigl | \hat c_{{\bf k}'_2\downarrow} \Bigr|n \Bigr\rangle\Biggr].
\nonumber \\ 
\label{eq:b1}
\end{eqnarray}
Using the integral representation of the step function,
\begin{equation}
\Theta(t)=\lim_{\delta\to 0^{+}}\int_{-\infty}^{\infty}\frac{d\omega}{2\pi i}\frac{e^{i\omega t}}{\omega-i\delta}
=-\lim_{\delta\to 0^{+}}\int_{-\infty}^{\infty}\frac{d\omega}{2\pi i}\frac{e^{-i\omega t}}{\omega+i\delta},
\end{equation}
and taking Fourier transform of (\ref{eq:b1}), we obtain  
\begin{eqnarray}
&&G_{\uparrow}({\bf k}_1,{\bf k}_2,\omega_1,\omega_2)
\nonumber \\
&&=\sum_n\sum_m\sum_l\sum_k \rho_n
\int_{-\infty}^{\infty}
\frac{d\omega}{2\pi i}
\frac{-1}{\omega + i\delta}
 \nonumber \\
&&~~\times \Biggl[
\frac{\bigl\langle n \bigl|\hat c^{\dagger}_{{\bf k}_2'\downarrow}\bigr |m \bigr\rangle 
\bigl\langle m \bigl |  \hat c^{\dagger}_{{\bf k}_1'\uparrow}\bigr|l\bigr\rangle
\bigl\langle l \bigl |\hat c_{{\bf k}_1'\uparrow }\bigr|k \bigr\rangle \bigl\langle k \bigl | \hat c_{{\bf k}_2'\downarrow}
\bigr|n \bigr\rangle}{\omega_1-\omega+E_m/\hbar-E_l/\hbar-i\eta}
 \nonumber \\
&&~~~~~~~~\times
\frac{1}{\omega_2+\omega+E_n/\hbar-E_m/\hbar-i\eta}
\frac{1}{E_k/\hbar-E_n/\hbar-\omega_2-i\eta}
\nonumber \\
&&~~~~~
-\frac{\bigl\langle n \bigl|\hat c^{\dagger}_{{\bf k}_1'\uparrow}\bigr|m\bigr\rangle
\bigl\langle m\bigl |\hat c^{\dagger}_{{\bf k}'_2\downarrow}\bigr|l \bigr\rangle \bigl\langle l \bigl |
\hat c_{{\bf k}'_1\uparrow } \bigr | k \bigr\rangle\bigl\langle k \bigl | \hat c_{{\bf k}'_2\downarrow} \Bigr|n \Bigr\rangle}
{\omega_1+\omega+E_n/\hbar-E_m/\hbar-i\eta }
 \nonumber \\
&&~~~~~~~~\times
\frac{1}{\omega_2-\omega+E_m/\hbar-E_l/\hbar-i\eta}
\frac{1}{E_k/\hbar-E_n/\hbar-\omega_2-i\eta}
\Biggr].
\nonumber \\ 
\end{eqnarray}
Taking the limit $\delta\to 0_+$ is implied in the above expression.
Carrying out the integral over $\omega$, we obtain
\begin{eqnarray}
&&G_{\uparrow}({\bf k}_1,{\bf k}_2,\omega_1,\omega_2)
=-\sum_n\sum_l\rho_n\nonumber \\
&&\times \Biggl\{\frac{1}{\omega_1+\omega_2-(E_l-E_n)/\hbar-2i\eta}\left|
\sum_m
\frac{\bigl\langle l \bigl |  \hat c_{{\bf k}_1'\uparrow}\bigr|m\bigr\rangle
\bigl\langle m \bigl|\hat c_{{\bf k}_2'\downarrow}\bigr |n \bigr\rangle
}
{\omega_2-(E_m-E_n)/\hbar+i\eta}\right|^2
\nonumber \\
&&~
-\frac{1}{\omega_2+\omega_1-(E_l-E_n)/\hbar -2i\eta}
\left[\sum_m
\frac{\bigl\langle l\bigl |\hat c_{{\bf k}'_2\downarrow}\bigr|m \bigr\rangle
\bigl\langle m \bigl|\hat c_{{\bf k}_1'\uparrow}\bigr|n\bigr\rangle
 }
{\omega_1-(E_m-E_n)/\hbar+i\eta}\right]^*
 \nonumber \\
&&~~~~~~~~\times
\left[\sum_m \frac{\bigl\langle l \bigl |
\hat c_{{\bf k}'_1\uparrow } \bigr | m \bigr\rangle\bigl\langle m \bigl | \hat c_{{\bf k}'_2\downarrow} \Bigr|n \Bigr\rangle}{\omega_2-(E_m-E_n)/\hbar+i\eta}
\right]\Biggr\}.
\end{eqnarray}

Similarly, we obtain  
\begin{eqnarray}
&&G_{\downarrow}({\bf k}_1,{\bf k}_2,\omega_1,\omega_2)
\nonumber \\
&&=-\sum_n\sum_l\rho_n \nonumber \\
&&\Biggl\{\frac{1}{\omega_1+\omega_2-(E_l-E_n)/\hbar-2i\eta}
\left|\sum_m\frac{\bigl\langle l \bigl|\hat c_{{\bf k}_2'\downarrow}\bigr|m\bigr\rangle
\bigl\langle m \bigl| \hat c_{{\bf k}_1'\uparrow}\bigr|n\bigr\rangle}
{\omega_1-(E_m-E_n)/\hbar+i\eta}\right|^2
\nonumber \\
&&-\frac{1}{\omega_2+\omega_1-(E_l-E_n)/\hbar-2i\eta}
\left[\sum_m\frac{\bigl\langle l\bigl|\hat c_{{\bf k}_1'\uparrow}\bigr|m\bigr\rangle
\bigl\langle m  \bigl| \hat c_{{\bf k}_2'\downarrow}\bigr|n\bigr\rangle
}{\omega_2-(E_m-E_n)/\hbar+i\eta}\right]^*
 \nonumber \\
&&~~~~~~~~\times
\left[\sum_m\frac{\bigl\langle l\bigl|\hat c_{{\bf k}_2'\downarrow }\bigr|m\bigr\rangle\bigl\langle m\bigl|\hat c_{{\bf k}_1'\uparrow} \bigr|n\bigr\rangle
}{\omega_1-(E_m-E_n)/\hbar+i\eta}\right]
\Biggr\}.
\end{eqnarray}

\section{Single-particle current}
The single-particle density in the scattered state is given by 
\begin{eqnarray}
n_\sigma({\bf k},t)=\langle \hat b^\dagger_{{\bf k}\sigma}\hat b_{\bf k\sigma}  \rangle_t.
\end{eqnarray}
Then, the single-particle current is given by 
\begin{eqnarray}
J_{\sigma}({\bf k},t)=\frac{d}{dt}n_{\sigma}({\bf k},t) = \frac{i}{\hbar} \langle[\hat b^\dagger_{\bf k\sigma}\hat b_{{\bf k}\sigma},\hat H ]\rangle_t.
\end{eqnarray}
With the Hamiltonin given by (\ref{eq:Hami}), we have
    \begin{eqnarray}
    J_{\sigma}({\bf k},t)=\frac{i}{\hbar}e^{\eta t}\left(\gamma_\sigma e^{-i\omega_\sigma t}
    \langle 
   \hat b^\dagger_{{\bf k} \sigma}  \hat c_{{\bf k}-{\bf q}_\sigma\sigma}
    \rangle_t
    - \gamma_\sigma^\ast e^{i\omega_\sigma t}
    \langle 
     \hat c^\dagger_{{\bf k}-{\bf q}_\sigma\sigma} \hat b_{{\bf k} \sigma}
    \rangle_t
     \right).
   \end{eqnarray}
   Employing the time-dependent perturbation theory, we can express the expectation values of $b^\dagger c$ and $c^\dagger b$ as
     \begin{eqnarray}
     \langle b^\dagger c \rangle =  \langle b^\dagger c \rangle^{(1)}+\langle b^\dagger c \rangle^{(3)}
    \end{eqnarray}
      \begin{eqnarray}
     \langle c^\dagger b \rangle =  \langle c^\dagger b \rangle^{(1)}+\langle c^\dagger b \rangle^{(3)}
    \end{eqnarray}
    where the superscript (n) denotes the n-th order contribution. The first order term gives the usual expression for the tunneling current, which is described in terms of the single-particle spectral function.
   We are now interested in the third-order contributions 
     \begin{eqnarray}
      \langle \hat b^\dagger_{{\bf k}\sigma } \hat c_{{\bf k}-{\bf q}_\sigma \sigma} \rangle_t^{(3)}&=&\frac{1}{6}\left(-\frac{i}{\hbar}\right)^3\int_c dt^\prime \int_c dt^{\prime\prime} \int_c dt^{\prime\prime\prime }\langle {\mathcal T}\left[
     \hat V(t^\prime)     \hat V(t^{\prime\prime})    \hat V(t^{\prime\prime\prime})
     \hat b^\dagger_{{\bf k}\sigma }(t) \hat c_{{\bf k}-{\bf q}_\sigma \sigma} (t)
      \right]\rangle_{t_0},
\\ 
      \langle  \hat c^\dagger_{{\bf k}-{\bf q}_\sigma \sigma} 
      \hat b_{{\bf k}\sigma }
      \rangle_t^{(3)}&=&\frac{1}{6}\left(-\frac{i}{\hbar}\right)^3\int_c dt^\prime \int_c dt^{\prime\prime} \int_c dt^{\prime\prime\prime }\langle {\mathcal T} \left[
     \hat V(t^\prime)     \hat V(t^{\prime\prime})    \hat V(t^{\prime\prime\prime})
     \hat b^\dagger_{{\bf k}\sigma }(t) \hat c_{{\bf k}-{\bf q}_\sigma \sigma} (t)
      \right]\rangle_{t_0}.
    \end{eqnarray}
    The non-vanishing contribution is 
    \begin{eqnarray}
    \langle\hat b^\dagger_{{\bf k} \sigma} \hat c_{{\bf k}-{\bf q}_\sigma \sigma}\rangle_t^{(3)}&&=\frac{i}{2\hbar^3}\sum_{\sigma, \sigma^\prime, \sigma^{\prime\prime}}\sum_{{\bf k}, {\bf k}^\prime, {\bf k}^{\prime\prime}}\gamma_{\sigma^\prime}^\ast\gamma_{\sigma^{\prime\prime}}^\ast\gamma_{\sigma^{\prime\prime\prime}}^\ast\int_c dt^\prime \int_c dt^{\prime\prime} \int_c dt^{\prime\prime\prime }e^{i(\omega_{\sigma^\prime}-i\eta)t^\prime}e^{i(\omega_{\sigma^{\prime\prime}}-i\eta)t^{\prime\prime}}e^{i(\omega_{\sigma^{\prime\prime\prime}}-i\eta)t^{\prime\prime\prime}}\nonumber \\ &&
    \langle {\mathcal T} [\hat c^\dagger_{{\bf k}^\prime \sigma^\prime}(t^\prime)
    \hat b_{{\bf k}^\prime +{\bf q}^\prime_{\sigma^\prime}\sigma^\prime}(t^\prime)
    \hat c^\dagger_{{\bf k}^{\prime\prime} \sigma^{\prime\prime}}(t^{\prime\prime})
    \hat b_{{\bf k}^{\prime\prime} +{\bf q}^{\prime\prime}_{\sigma^{\prime\prime}}\sigma^{\prime\prime}}(t^{\prime\prime})
     \nonumber \\
&&~~~~~~~~\times
    \hat b^\dagger_{{\bf k}^{\prime\prime\prime} +{\bf q}^{\prime\prime\prime}_{\sigma^{\prime\prime\prime}}\sigma^{\prime\prime\prime}}(t^{\prime\prime\prime})
    \hat c_{{\bf k}^{\prime\prime\prime} \sigma^{\prime\prime\prime}}(t^{\prime\prime\prime})
    \hat b^\dagger_{{\bf k}\sigma}(t)
     \hat c_{{\bf k}  -{\bf q}_{\sigma}\sigma}(t)
     ]\rangle_{t_0}.
      \label{eq:c8}
    \end{eqnarray}
    With the assumption that the two internal states are initially uncoupled and the scattered state is a free gas, the above correlation function can be decoupled as 
     \begin{eqnarray}
     && \langle {\mathcal T} [\hat c^\dagger_{{\bf k}^\prime \sigma^\prime}(t^\prime)
    \hat b_{{\bf k}^\prime +{\bf q}^\prime_{\sigma^\prime}\sigma^\prime}(t^\prime)
    \hat c^\dagger_{{\bf k}^{\prime\prime} \sigma^{\prime\prime}}(t^{\prime\prime})
    \hat b_{{\bf k}^{\prime\prime} +{\bf q}^{\prime\prime}_{\sigma^{\prime\prime}}\sigma^{\prime\prime}}(t^{\prime\prime})
    \hat b^\dagger_{{\bf k}^{\prime\prime\prime} +{\bf q}^{\prime\prime\prime}_{\sigma^{\prime\prime\prime}}\sigma^{\prime\prime\prime}}(t^{\prime\prime\prime})
    \hat c_{{\bf k}^{\prime\prime\prime} \sigma^{\prime\prime\prime}}(t^{\prime\prime\prime})
    \hat b^\dagger_{{\bf k} \sigma}(t)
     \hat c_{{\bf k} -{\bf q}_{\sigma}\sigma}(t)
     ]\rangle_{t_0}
     \nonumber \\
    && =
    \langle {\mathcal T} [\hat c^\dagger_{{\bf k}^\prime \sigma^\prime}(t^\prime)
    \hat c^\dagger_{{\bf k}^{\prime\prime} \sigma^{\prime\prime}}(t^{\prime\prime})
        \hat c_{{\bf k}^{\prime\prime\prime} \sigma^{\prime\prime\prime}}(t^{\prime\prime\prime})
     \hat c_{{\bf k}-{\bf q}_{\sigma} \sigma}(t)
           ]\rangle_{t_0}
            \nonumber \\
&&~~~~~~\times
           \langle {\mathcal T} [
     \hat b_{{\bf k}^\prime +{\bf q}^\prime_{\sigma^\prime}\sigma^\prime}(t^\prime)
    \hat b_{{\bf k}^{\prime\prime} +{\bf q}^{\prime\prime}_{\sigma^{\prime\prime}}\sigma^{\prime\prime}}(t^{\prime\prime})
    \hat b^\dagger_{{\bf k}^{\prime\prime\prime} +{\bf q}^{\prime\prime\prime}_{\sigma^{\prime\prime\prime}}\sigma^{\prime\prime\prime}}(t^{\prime\prime\prime})
     \hat b^\dagger_{{\bf k} \sigma}(t)
     ]\rangle_{t_0}
\nonumber \\ 
&&=
   - \Theta(t^\prime,t^{\prime\prime\prime})\Theta(t^{\prime\prime},t)
   \delta_{{\bf k}^{\prime\prime}+{{\bf q}_\sigma^{\prime\prime}},{\bf k}}\delta_{\sigma^{\prime\prime},\sigma^\prime}\delta_{{\bf k}^{\prime}+{{\bf q}_\sigma^{\prime}},{\bf k}^{\prime\prime\prime}+{{\bf q}_\sigma^{\prime\prime\prime}}}\delta_{\sigma^{\prime\prime}\sigma^{\prime\prime\prime}}e^{-i\epsilon_{{\bf k} \sigma}(t^\prime-t)}e^{-i\epsilon_{{\bf k}^{\prime\prime}+{\bf q}_{\sigma}^{\prime\prime}\sigma^{\prime\prime}}(t^{\prime}-t^{\prime\prime\prime})}.
   \label{eq:c9}
      \end{eqnarray}  
Using (\ref{eq:c8}) and (\ref{eq:c9}), we obtain 
\begin{eqnarray}
&&\langle b^\dagger_{{\bf k} \sigma} c_{{\bf k}-{\bf q}_\sigma \sigma}\rangle_t^{(3)}\nonumber \\
&&=\frac{i}{\hbar^3} \sum_{\sigma^{\prime}}\sum_{{\bf k}^{\prime}}\gamma_\sigma^\ast |\gamma_{\sigma^{\prime}}|^2 \int_c dt^\prime  \int_c dt^{\prime\prime}
\nonumber \\
&&\Theta(t^{\prime\prime}, t)
\Theta(t^\prime, t^{\prime\prime\prime})e^{-i\epsilon_{{\bf k}\sigma}(t^{\prime\prime}-t)/\hbar}e^{-i\epsilon_{{\bf k}^{\prime}+{\bf q}_{ \sigma^{\prime}} \sigma^{\prime}}(t^{\prime}-t^{\prime\prime\prime})/\hbar} \nonumber \\
&&~~~~~~~~\times
\langle {\mathcal T} \left[
\hat c^\dagger_{{\bf k}-{\bf q}_{\sigma}\sigma}(t^{\prime\prime})
\hat c^\dagger_{{\bf k}^{\prime}\sigma^{\prime}}(t^\prime) 
\hat c_{{\bf k}^{\prime}\sigma^{\prime}}(t^{\prime\prime\prime})
\hat c_{{\bf k}-{\bf q}_{\sigma}\sigma}(t)
\right]\rangle_{t_0}.\label{eq:c10}
  \end{eqnarray} 
  In order to express (\ref{eq:c10}) in terms of the real-time integral, we split the contour integral into the forward (\overleftarrow{c}) and return  (\overleftarrow{c}) paths. Then the contour integral involving the product of step functions can be written as
\begin{eqnarray}
&&\int_c dt^\prime  \int_c dt^{\prime\prime} \int_c dt^{\prime\prime\prime} \Theta (t^{\prime\prime},t)\Theta (t^{\prime},t^{\prime\prime\prime})
 \nonumber \\&&=
 \int_{\overleftarrow{c}} dt^{\prime\prime}\Bigg[\int_{\overrightarrow{c}} dt^\prime\int_{\overrightarrow{c}} dt^{\prime\prime\prime}\theta(t^\prime-t^{\prime\prime\prime})
 \nonumber \\&&
 +
 \int_{\overleftarrow{c}} dt^\prime\int_{\overrightarrow{c}} dt^{\prime\prime\prime}
 +\int_{\overleftarrow{c}} dt^\prime\int_{\overleftarrow{c}} dt^{\prime\prime\prime}\theta(t^{\prime\prime\prime}-t^\prime)
 \Bigg].
\end{eqnarray} \label{eq:c11}
\begin{figure}[htbp]
\includegraphics[width=120mm]{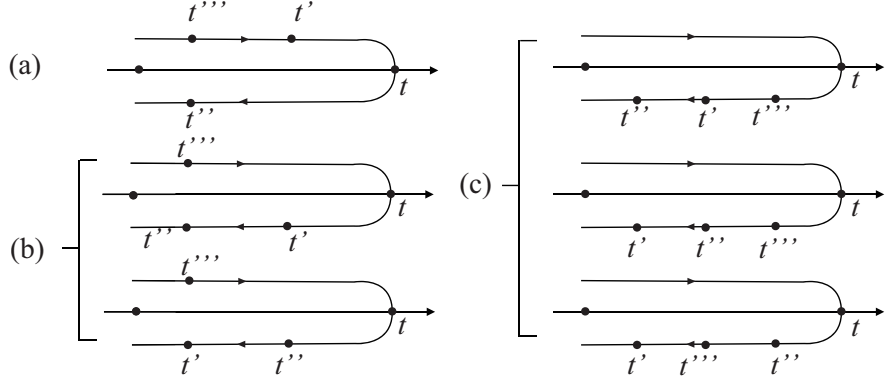}
\caption{\label{fig_3} The counter path split into forward and return branches}
\end{figure}
Figure 6 depicts the three contributions in the square bracket of (\ref{eq:c11})
We now calculate the contribution depicted by Fig.~\ref{fig_3} (b), which is given by 
\begin{eqnarray}
&& \int_{\vec{c}} dt^{\prime\prime} \int_{\vec{c}} dt^\prime\int_{\vec{c}} dt^{\prime\prime\prime}
 e^{i(\omega_\sigma-i\eta)t^{\prime\prime}}e^{i(\omega_{\sigma^{\prime}}-i\eta)(t^{\prime}-t^{\prime\prime\prime})}e^{-i\epsilon_{{\bf k}\sigma}(t^{\prime\prime}-t)/\hbar}
 e^{-i\epsilon_{{\bf k}^{\prime}+{\bf q}_{ \sigma^{\prime}} \sigma^{\prime}}(t^{\prime}-t^{\prime\prime\prime})/\hbar}
  \nonumber \\&&
\langle {\mathcal T} \left[
\hat c^\dagger_{{\bf k}-{\bf q}_{\sigma}\sigma}(t^{\prime\prime})
\hat c^\dagger_{{\bf k}^{\prime}\sigma^{\prime}}(t^\prime) 
\hat c_{{\bf k}^{\prime}\sigma^{\prime}}(t^{\prime\prime\prime})
\hat c_{{\bf k}-{\bf q}_{\sigma}\sigma}(t)
\right]\rangle_{t_0}
\nonumber \\&&=e^{i(\omega_\sigma-i\eta)t}
 \int_{t_0}^0dt_1 \int_{t_0}^0dt_2 \int_{t_0}^0dt_3
 e^{i(\omega_{\sigma^{\prime}}-\epsilon_{{\bf k}^{\prime}+{\bf q}_{ \sigma^{\prime}} \sigma^{\prime}}/\hbar-i\eta)t_1}
 e^{i(\omega_\sigma-\epsilon_{{\bf k}\sigma}/\hbar-i\eta)t_2}
  e^{-i(\omega_{\sigma^{\prime}}-\epsilon_{{\bf k}^{\prime}+{\bf q}_{ \sigma^{\prime}} \sigma^{\prime}}/\hbar+i\eta)t_3}
  \nonumber \\&&
      \langle \tilde{\mathcal T} \left[
  \hat c^\dagger_{{\bf k}^{\prime}\sigma^{\prime}}(t_1) 
\hat c^\dagger_{{\bf k}-{\bf q}_{\sigma}\sigma}(t_2)
\right]
\hat c_{{\bf k}-{\bf q}_{\sigma}\sigma}(0)
\hat c_{{\bf k}^{\prime}\sigma^{\prime}}(t_3)
\rangle_{t_0}.
\end{eqnarray} 
Here we have introduced the relative time coordinates
\begin{equation}
t_1=t'-t,~~t_2=t''-t,~~t_3=t'''-t.
\end{equation}
Using the notations
$k-q_\sigma=k_\sigma$, $\omega_{\sigma^\prime}^{\prime\prime}=\omega_{\sigma^{\prime}}-\epsilon_{{\bf k}^{\prime}+{\bf q}_{ \sigma^{\prime}}}/\hbar$, $\omega_\sigma^\prime=\omega_\sigma-\epsilon_{{\bf k}\sigma}/\hbar$,
we obtain the expression for the
contribution Fig.~\ref{fig_3} (b) to the
single-particle current as
\begin{eqnarray}
J_{\sigma}^{(b)}=-\frac{1}{\hbar^4}\sum_{\sigma^\prime}\sum_{k^\prime}|\gamma_\sigma|^2|\gamma_{\sigma^\prime}|^2
G_{\sigma\sigma^\prime}(k^\prime,k_\sigma,\omega_{\sigma^{\prime}}^{\prime\prime}-i\eta,\omega_\sigma^\prime-i\eta,\omega_{\sigma^{\prime}}^{\prime\prime}+i\eta),
\end{eqnarray}
where we defined the Fourier transforms by   
\begin{eqnarray}
G_{\sigma\sigma^\prime}(k_1,k_2,\omega_{1\sigma'},\omega_{2\sigma},\omega_{3\sigma'})
=\int_{-\infty}^{\infty}dt_1\int_{-\infty}^{\infty}dt_2\int_{-\infty}^{\infty}dt_3e^{i\omega_{1\sigma'}t_1}e^{i\omega_{2\sigma}t_2}e^{-i\omega_3t_3}
G_{\sigma\sigma^\prime}(k_1,k_2,t_1,t_2,t_3),
\nonumber \\ 
\end{eqnarray}
with 
\begin{eqnarray}
iG_{\sigma\sigma^\prime}(k^\prime,k_\sigma,t_1,t_2,t_3)
=      \langle \tilde{\mathcal T} \left[
  \hat c^\dagger_{{\bf k}^{\prime}\sigma^{\prime}}(t_1) 
\hat c^\dagger_{{\bf k}-{\bf q}_{\sigma}\sigma}(t_2)
\right]
\hat c_{{\bf k}-{\bf q}_{\sigma}\sigma}(0)
\hat c_{{\bf k}^{\prime}\sigma^{\prime}}(t_3)
\rangle_{t_0}\theta (-t_1)\theta (-t_2)\theta (-t_3)
\nonumber \\ 
\end{eqnarray}

The contribution of Fig.~\ref{fig_3} (a) is given by 
\begin{eqnarray}
&& \int_{\vec{c}} dt^{\prime\prime} \int_{\vec{c}} dt^\prime\int_{\vec{c}} dt^{\prime\prime\prime}
\theta (t{'}-t{'''})
 e^{i(\omega_\sigma-i\eta)t^{\prime\prime}}e^{i(\omega_{\sigma^{\prime}}-i\eta)(t^{\prime}-t^{\prime\prime\prime})}e^{-i\epsilon_{{\bf k}\sigma}(t^{\prime\prime}-t)/\hbar}
 e^{-i\epsilon_{{\bf k}^{\prime}+{\bf q}_{ \sigma^{\prime}} \sigma^{\prime}}(t^{\prime}-t^{\prime\prime\prime})/\hbar}
  \nonumber \\&&
\langle {\mathcal T} \left[
\hat c^\dagger_{{\bf k}-{\bf q}_{\sigma}\sigma}(t^{\prime\prime})
\hat c^\dagger_{{\bf k}^{\prime}\sigma^{\prime}}(t^\prime) 
\hat c_{{\bf k}^{\prime}\sigma^{\prime}}(t^{\prime\prime\prime})
\hat c_{{\bf k}-{\bf q}_{\sigma}\sigma}(t)
\right]\rangle_{t_0}
\nonumber \\ &&
=-e^{i(\omega_\sigma-i\eta)t}
 \int_{t_0}^0dt_1 \int_{t_0}^0dt_2 \int_{t_0}^0dt_3
 \theta (t_1-t_3)
  \nonumber \\
&&~~~~~~~~\times
 e^{i(\omega_{\sigma^{\prime}}-\epsilon_{{\bf k}^{\prime}+{\bf q}_{ \sigma^{\prime}} \sigma^{\prime}}/\hbar-i\eta)t_1}
 e^{i(\omega_\sigma-\epsilon_{{\bf k}\sigma}/\hbar-i\eta)t_2}
  e^{-i(\omega_{\sigma^{\prime}}-\epsilon_{{\bf k}^{\prime}+{\bf q}_{ \sigma^{\prime}} \sigma^{\prime}}/\hbar+i\eta)t_3}
  \nonumber \\&&~~~~~~~~\times
  \langle
\hat c^\dagger_{{\bf k}-{\bf q}_{\sigma}\sigma}(t_2)
\hat c_{{\bf k}-{\bf q}_{\sigma}\sigma}(0)
\hat c^\dagger_{{\bf k}^{\prime}\sigma^{\prime}}(t_1) 
\hat c_{{\bf k}^{\prime}\sigma^{\prime}}(t_3)
\rangle_{t_0}.
\end{eqnarray} 
Therefore we obtain its contribution to the single-particle current as
\begin{eqnarray}
J_{\sigma}^{(a)}=-\frac{1}{\hbar^4}\sum_{\sigma^\prime}\sum_{k^\prime}|\gamma_\sigma|^2|\gamma_{\sigma^\prime}|^2
G'_{\sigma\sigma^\prime}(k^\prime,k_\sigma,\omega_{\sigma^{\prime}}^{\prime\prime}-i\eta,\omega_\sigma^\prime-i\eta,\omega_{\sigma^{\prime}}^{\prime\prime}+i\eta),
\end{eqnarray}
where 
\begin{eqnarray}
G'_{\sigma\sigma^\prime}(k_1,k_2,\omega_{1\sigma'},\omega_{2\sigma},\omega_3)
=\int_{-\infty}^{\infty}dt_1\int_{-\infty}^{\infty}dt_2\int_{-\infty}^{\infty}dt_3e^{i\omega_{1\sigma'}t_1}e^{i\omega_{2\sigma}t_2}e^{-i\omega_3t_3}
G'_{\sigma\sigma^\prime}(k_1,k_2,t_1,t_2,t_3), 
\nonumber \\ 
\end{eqnarray}
and the correlation function is given by 
\begin{eqnarray}
 &&iG'_{\sigma\sigma^\prime}(k^\prime,k_\sigma,t_1,t_2,t_3)
 =-\langle
\hat c^\dagger_{{\bf k}_{\sigma}\sigma}(t_2)
\hat c_{{\bf k}_{\sigma}\sigma}(0)
\hat c^\dagger_{{\bf k}^{\prime}\sigma^{\prime}}(t_1) 
\hat c_{{\bf k}^{\prime}\sigma^{\prime}}(t_3)
\rangle_{t_0}
\theta (t_1-t_3)  \theta (-t_1)\theta (-t_2)\theta (-t_3).
\nonumber \\ 
\end{eqnarray} 

The contribution of Fig.~\ref{fig_3} (c) is given by 
\begin{eqnarray}
&& \int_{\vec{c}} dt^{\prime\prime} \int_{\vec{c}} dt^\prime\int_{\vec{c}} dt^{\prime\prime\prime}
\theta (t{'''}-t{'})
 e^{i(\omega_\sigma-i\eta)t^{\prime\prime}}e^{i(\omega_{\sigma^{\prime}}-i\eta)(t^{\prime}-t^{\prime\prime\prime})}e^{-i\epsilon_{{\bf k}\sigma}(t^{\prime\prime}-t)/\hbar}
 e^{-i\epsilon_{{\bf k}^{\prime}+{\bf q}_{ \sigma^{\prime}} \sigma^{\prime}}(t^{\prime}-t^{\prime\prime\prime})/\hbar}
  \nonumber \\&&
\langle {\mathcal T} \left[
\hat c^\dagger_{{\bf k}-{\bf q}_{\sigma}\sigma}(t^{\prime\prime})
\hat c^\dagger_{{\bf k}^{\prime}\sigma^{\prime}}(t^\prime) 
\hat c_{{\bf k}^{\prime}\sigma^{\prime}}(t^{\prime\prime\prime})
\hat c_{{\bf k}-{\bf q}_{\sigma}\sigma}(t)
\right]\rangle_{t_0}
  \nonumber \\&&
=-e^{i(\omega_\sigma-i\eta)t}
 \int_{t_0}^0dt_1 \int_{t_0}^0dt_2 \int_{t_0}^0dt_3
 e^{i(\omega_{\sigma^{\prime}}-\epsilon_{{\bf k}^{\prime}+{\bf q}_{ \sigma^{\prime}} \sigma^{\prime}}/\hbar-i\eta)t_1}
 e^{i(\omega_\sigma-\epsilon_{{\bf k}\sigma}/\hbar-i\eta)t_2}
  e^{-i(\omega_{\sigma^{\prime}}-\epsilon_{{\bf k}^{\prime}+{\bf q}_{ \sigma^{\prime}} \sigma^{\prime}}/\hbar+i\eta)t_3}
  \nonumber \\&&
   \theta (t_3-t_1)
    \Bigl\{
-   \theta (t_1-t_2)\langle
\hat c^\dagger_{{\bf k}-{\bf q}_{\sigma}\sigma}(t_2)
\hat c^\dagger_{{\bf k}^{\prime}\sigma^{\prime}}(t_1)
\hat c_{{\bf k}^{\prime}\sigma^{\prime}}(t_3)
\hat c_{{\bf k}-{\bf q}_{\sigma}\sigma}(0)\rangle_{t_0}
 \nonumber \\&&~~~~~~~~~~~~~
+   \theta (t_3-t_2)   \theta (t_2-t_1)\langle
\hat c^\dagger_{{\bf k}^{\prime}\sigma^{\prime}}(t_1)
\hat c^\dagger_{{\bf k}-{\bf q}_{\sigma}\sigma}(t_2)
\hat c_{{\bf k}^{\prime}\sigma^{\prime}}(t_3)
\hat c_{{\bf k}-{\bf q}_{\sigma}\sigma}(0)\rangle_{t_0}
 \nonumber \\&&~~~~~~~~~~~~~
-   \theta (t_2-t_3)\langle
\hat c^\dagger_{{\bf k}^{\prime}\sigma^{\prime}}(t_1)
\hat c_{{\bf k}^{\prime}\sigma^{\prime}}(t_3)
\hat c^\dagger_{{\bf k}-{\bf q}_{\sigma}\sigma}(t_2)
\hat c_{{\bf k}-{\bf q}_{\sigma}\sigma}(0)\rangle_{t_0}
  \Bigl\}.
\end{eqnarray}
Therefore, we obtain its contribution to the single-particle current as
\begin{eqnarray}
J_{\sigma}^{(c)}=-\frac{1}{\hbar^4}\sum_{\sigma^\prime}\sum_{k^\prime}|\gamma_\sigma|^2|\gamma_{\sigma^\prime}|^2
G''_{\sigma\sigma^\prime}(k^\prime,k_\sigma,\omega_{\sigma^{\prime}}^{\prime\prime}-i\eta,\omega_\sigma^\prime-i\eta,\omega_{\sigma^{\prime}}^{\prime\prime}+i\eta),
\end{eqnarray}

where we defined  
\begin{eqnarray}
&&G''_{\sigma\sigma^\prime}(k_1,k_2,\omega_{1\sigma'},\omega_{2\sigma},\omega_3)
  \nonumber \\ &&
=\int_{-\infty}^{\infty}dt_1\int_{-\infty}^{\infty}dt_2\int_{-\infty}^{\infty}dt_3e^{i\omega_{1\sigma'}t_1}e^{i\omega_{2\sigma}t_2}e^{-i\omega_3t_3}
G''_{\sigma\sigma^\prime}(k_1,k_2,t_1,t_2,t_3),
\end{eqnarray}
with 
\begin{eqnarray}
 &&iG''_{\sigma\sigma^\prime}(k^\prime,k_\sigma,t_1,t_2,t_3)
  \nonumber \\&&
   = 
  \theta(-t_1)\theta(-t_2)\theta(-t_3) \theta (t_3-t_1)
    \Bigl\{
-   \theta (t_1-t_2)\langle
\hat c^\dagger_{{\bf k}-{\bf q}_{\sigma}\sigma}(t_2)
\hat c^\dagger_{{\bf k}^{\prime}\sigma^{\prime}}(t_1)
\hat c_{{\bf k}^{\prime}\sigma^{\prime}}(t_3)
\hat c_{{\bf k}-{\bf q}_{\sigma}\sigma}(0)\rangle_{t_0}
 \nonumber \\&&
+   \theta (t_3-t_2)   \theta (t_2-t_1)\langle
\hat c^\dagger_{{\bf k}^{\prime}\sigma^{\prime}}(t_1)
\hat c^\dagger_{{\bf k}-{\bf q}_{\sigma}\sigma}(t_2)
\hat c_{{\bf k}^{\prime}\sigma^{\prime}}(t_3)
\hat c_{{\bf k}-{\bf q}_{\sigma}\sigma}(0)\rangle_{t_0}
 \nonumber \\&&
-   \theta (t_2-t_3)\langle
\hat c^\dagger_{{\bf k}^{\prime}\sigma^{\prime}}(t_1)
\hat c_{{\bf k}^{\prime}\sigma^{\prime}}(t_3)
\hat c^\dagger_{{\bf k}-{\bf q}_{\sigma}\sigma}(t_2)
\hat c_{{\bf k}-{\bf q}_{\sigma}\sigma}(0)\rangle_{t_0}
  \Bigl\}.
\end{eqnarray}
Collecting the above results, we obtain the expression for the third-order contribution to the singleparticle
current as
\begin{eqnarray}
&&J_{\sigma}({\bf k},t)=
J_{\sigma}^{(a)}+J_{\sigma}^{(c)}+J_{\sigma}^{(c)}\nonumber \\
&&
=-\frac{1}{\hbar^4}\sum_{\sigma^\prime}\sum_{k^\prime}|\gamma_\sigma|^2|\gamma_{\sigma^\prime}|^2
G_{\sigma\sigma^\prime}(k^\prime,k_\sigma,\omega_{\sigma^{\prime}}^{\prime\prime}-i\eta,\omega_\sigma^\prime-i\eta,\omega_{\sigma^{\prime}}^{\prime\prime}+i\eta)
\nonumber \\
&&
-\frac{1}{\hbar^4}\sum_{\sigma^\prime}\sum_{k^\prime}|\gamma_\sigma|^2|\gamma_{\sigma^\prime}|^2
G'_{\sigma\sigma^\prime}(k^\prime,k_\sigma,\omega_{\sigma^{\prime}}^{\prime\prime}-i\eta,\omega_\sigma^\prime-i\eta,\omega_{\sigma^{\prime}}^{\prime\prime}+i\eta)
\nonumber \\
&&
-\frac{1}{\hbar^4}\sum_{\sigma^\prime}\sum_{k^\prime}|\gamma_\sigma|^2|\gamma_{\sigma^\prime}|^2
G''_{\sigma\sigma^\prime}(k^\prime,k_\sigma,\omega_{\sigma^{\prime}}^{\prime\prime}-i\eta,\omega_\sigma^\prime-i\eta,\omega_{\sigma^{\prime}}^{\prime\prime}+i\eta).
\end{eqnarray}

Using the energy representation introduced in Sec III, we can express the two-particle correction function.
Taking the limit $\delta\to 0^+$, we thus have 
\begin{eqnarray}
&&G_{\sigma\sigma^\prime}(k^\prime,k-{\bf q}_{\sigma},\omega_{1\sigma'},\omega_{2\sigma},\omega_{1\sigma'})
\nonumber \\ && 
=\sum_{n,m,l,k}\rho_n \int \frac{d\omega}{2\pi }\frac{1}{\omega+i\delta}\int_{-\infty}^{0}dt_1\int_{-\infty}^{0}dt_2\int_{-\infty}^{0}dt_3
e^{i(\omega_{1\sigma'}-i\eta)t_1}e^{i(\omega_{2\sigma}-i\eta)t_2}e^{-i(\omega_{1\sigma'}-i\eta)t_3}
\nonumber \\ &&
\Big[e^{-i\omega(t_2-t_1)}e^{i(E_n-E_m)t_2/\hbar}e^{i(E_m-E_l)t_1/\hbar}e^{i(E_k-E_n)t_3/\hbar}
\langle n | \hat c^\dagger_{{\bf k}^{\prime}\sigma^{\prime}} |m \rangle 
\langle m | \hat c^\dagger_{{\bf k}-{\bf q}_{\sigma}\sigma}|l\rangle 
\langle l | \hat c_{{\bf k}-{\bf q}_{\sigma}\sigma} |k \rangle 
\langle k | \hat c_{{\bf k}^{\prime}\sigma^{\prime}}|n \rangle 
\nonumber \\ &&
-e^{-i\omega(t_1-t_2)}e^{i(E_n-E_m)t_1/\hbar}e^{i(E_m-E_l)t_2/\hbar}e^{i(E_k-E_n)t_3/\hbar}
\langle n | \hat c^\dagger_{{\bf k}-{\bf q}_{\sigma}\sigma} |m \rangle 
\langle m |\hat c^\dagger_{{\bf k}^{\prime}\sigma^{\prime}} |l\rangle 
\langle l | \hat c_{{\bf k}-{\bf q}_{\sigma}\sigma} |k \rangle 
\langle k | \hat c_{{\bf k}^{\prime}\sigma^{\prime}}|n \rangle\Big ]
\nonumber \\ &&
=-\sum_{n,l}\rho_n
\frac{1}{\omega_{1\sigma'}+\omega_{2\sigma}+(E_n-E_l)/\hbar-2i\eta}
\Bigg\{\Bigg|\sum_m\frac{\langle l | \hat c_{{\bf k}-{\bf q}_{\sigma}\sigma} |m\rangle 
\langle m | \hat c_{{\bf k}^{\prime}\sigma^{\prime}}|n \rangle }
{\omega_{1\sigma'}+(E_m-E_n)/\hbar+i\eta}\Bigg|^2
\nonumber \\ &&
-\left[\sum_m
\frac{\langle n | \hat c_{{\bf k}^{\prime}\sigma^{\prime}} 
 |m \rangle 
\langle m |\hat c_{{\bf k}-{\bf q}_{\sigma}\sigma}|l\rangle }
{\omega_{2\sigma}+(E_m-E_l)/\hbar-i\eta}\right]^\ast
\left[\sum_m
\frac{\langle l | \hat c_{{\bf k}-{\bf q}_{\sigma}\sigma} |m \rangle 
\langle m | \hat c_{{\bf k}^{\prime}\sigma^{\prime}}|n \rangle}
{\omega_{1\sigma'}+(E_m-E_n)/\hbar+i\eta}\right]
\Bigg\}.
\end{eqnarray}
Similarly, we obtain 
\begin{eqnarray}
&&G'_{\sigma\sigma^\prime}(k^\prime,k-{\bf q}_{\sigma},\omega_{1\sigma'},\omega_{2\sigma},\omega_{1\sigma'})
\nonumber \\ &&
=\sum_{n,m,l,k}\rho_n
\frac{\langle n |
\hat c^\dagger_{{\bf k}-{\bf q}_{\sigma}\sigma}|m\rangle 
\langle m |\hat c_{{\bf k}-{\bf q}_{\sigma}\sigma}|l\rangle 
\langle l |\hat c^\dagger_{{\bf k}^{\prime}\sigma^{\prime}} |k\rangle 
\langle k |\hat c_{{\bf k}^{\prime}\sigma^{\prime}}
|n\rangle }{\omega_{1\sigma'}-(E_k-E_n)/\hbar-i\eta}
\nonumber \\ && \times 
\frac{1}{\omega_{2\sigma}+(E_n-E_m)/\hbar-i\eta}
\frac{1}{(E_l+E_n-2E_k)/\hbar-2i\eta},
\nonumber \\ &&
\end{eqnarray}
and 
\begin{eqnarray}
&&G''_{\sigma\sigma^\prime}(k^\prime,k-{\bf q}_{\sigma},\omega_{1\sigma'},\omega_{2\sigma},\omega_{1\sigma'})
  \nonumber \\ &&
=-\sum_{n,m,l,k} \rho_n 
    \Biggl\{\frac{\langle n |\hat c^\dagger_{{\bf k}-{\bf q}_{\sigma}\sigma}|m\rangle
\langle m |\hat c^\dagger_{{\bf k}^{\prime}\sigma^{\prime}}|l\rangle
}{\omega_{1\sigma'}- (E_l-E_k)/\hbar-i\eta}\frac{\langle l |\hat c_{{\bf k}^{\prime}\sigma^{\prime}}|k\rangle
\langle k |\hat c_{{\bf k}-{\bf q}_{\sigma}\sigma}|n\rangle}{\omega_{2\sigma}+(E_n-E_k)/\hbar-i\eta}
\frac{1}{(E_k-E_m)/\hbar}
 \nonumber \\&&
+ 
\frac{\langle n|\hat c^\dagger_{{\bf k}^{\prime}\sigma^{\prime}}|m\rangle
\langle m |\hat c^\dagger_{{\bf k}-{\bf q}_{\sigma}\sigma}|l\rangle
}{\omega_{1\sigma'}+ (E_n-E_m)/\hbar-i\eta}
\frac{\langle l |\hat c_{{\bf k}^{\prime}\sigma^{\prime}} |k\rangle
\langle k |\hat c_{{\bf k}-{\bf q}_{\sigma}\sigma}|n\rangle}{\omega_{2\sigma}+(E_m-E_l)/\hbar-i\eta }
\frac{1}{\omega_{1\sigma'}-\omega_{2\sigma}+(E_k-E_m)/\hbar-i\eta}
 \nonumber \\&&
+ 
\frac{\langle n|\hat c^\dagger_{{\bf k}^{\prime}\sigma^{\prime}}|m\rangle
\langle m |\hat c_{{\bf k}^{\prime}\sigma^{\prime}}|k\rangle
}{-\omega_{1\sigma'}- (E_n-E_m)/\hbar+i\eta }
\frac{\langle k |\hat c^\dagger_{{\bf k}-{\bf q}_{\sigma}\sigma}|l\rangle
\langle l |\hat c_{{\bf k}-{\bf q}_{\sigma}\sigma}|n\rangle}{-\omega_{2\sigma}-(E_k-E_l)/\hbar+i\eta }
\frac{1}{-\omega_{1\sigma'}+\omega_{2\sigma}+ (E_m-E_l)/\hbar }
  \Biggl\}.
  \nonumber \\ 
\end{eqnarray}
Using the above results, we can obtain Eq.~(\ref{Eq_s}).


\bibliographystyle{apsrev}

\end{document}